\documentclass[12pt]{ar}
\usepackage[english]{babel}
\usepackage{psfig,graphicx}
\usepackage{supertab}

\onecolumn
\begin{document}

\title{V2324\,Cyg -- an F-type star with fast wind}

\author{V.G.\,Klochkova, E.L.\,Chentsov \& V.E.\,Panchuk}

\date{\today}	     

\institute{Special Astrophysical Observatory RAS, Nizhnij Arkhyz,  369167 Russia} 

\abstract{For the first time high-resolution optical spectroscopy of the
variable star V2324\,Cyg associated with the IR-source IRAS\,20572+4919 is
made. More than 200 absorption features (mostly Fe\,II, Ti\,II, Cr\,II,
Y\,II, Ba\,II, and Y\,II) are identified within the wavelength interval
4549--7880\,\AA{}. The spectral type and rotation velocity of the star are
found to be F0\,III and $V\sin i=69\,\mbox{km/s}$, respectively. H\,I and
NaI~D lines have complex P\,Cyg-type profiles with an emission component.
Neither systematic trend of radial velocity ${\rm V_r}$ with line depth
R${\rm _o}$ nor temporal variability of ${\rm V_r}$ have been found. We
determined the average heliocentric radial velocity ${\rm V_r}$=$-16.8 \pm
0.6$\,km/s. The radial velocities inferred from the cores of the
absorption components of the H$\beta$ and Na\,I wind lines vary from
$-140$ to $-225$\,km/s (and the expansion velocities of the corresponding
layers, from about 120 to 210\,km/s). The maximum expansion velocity is
found for the blue component of the split H$\alpha$ absorption: 450\,km/s
for December 12, 1995. The model atmospheres method is used to determine
the star's parameters: $T_{eff}$\,=7500\,K, $\log g$\,=\,2.0,
$\xi_t$\,=\,6.0\,km/s, and metallicity, which is equal to the solar value.
The main peculiarity of the chemical abundances pattern is the
overabundance of lithium and sodium. The results cast some doubt on the
classification of V2324\,Cyg as a post-AGB star. }

\titlerunning{\bf V2324\,Cyg---an F-type star with fast wind}
\authorrunning{\bf Klochkova et al.}

\maketitle

\section{Introduction}

The poorly studied star V2324\,Cyg (visible magnitude
V=11$\lefteqn{.}^{\rm m}63$, color indices B$-$V=+1$\lefteqn{.}^{\rm m}09$
and U$-$B=+0$\lefteqn{.}^{\rm m}58$, and galactic coordinates
l=89$\lefteqn{.}^{\rm o}44$, b=2$\lefteqn{.}^{\rm o}39$) is an optical
counterpart of the IR-source IRAS\,20572+4919. The observed
12--60\,$\mu\mbox{m}$ flux of the star and its location on the IR
color-color diagram suggest that this object is a candidate young
planetary nebula with a dust envelope [\cite{Garcia-1990}].

According to current theories (see, e.g., [\cite{Block}])
objects observed during the short-lived evolutionary stage of young
planetary (protoplanetary) nebulae (pPN) are intermediate-mass stars
evolving off the asymptotic giant branch (AGB) toward the planetary nebula
phase. The initial mass of these stars lies in the interval
3--8\,${\mathcal M}_{\odot}$. During the preceding AGB stage these stars
have undergone essential mass loss in the form of powerful stellar wind,
so that the resulting protoplanetary nebula has the form of a degenerate
carbon--oxygen core with a typical mass of about 0.6${\mathcal M}_{\odot}$
surrounded by an expanding gas-and-dust envelope. The astronomers'
interest toward pPN is due, first, to the fact that these objects allow us
to study the history of mass loss  and, second, because
they offer us an unique opportunity to observe the result of stellar
nucleosynthesis, mixing, and dredge-up of the products of nuclear
reactions during the previous evolution of the star.

The particular attention of the astronomers towards pPN over the past
decade yielded a number of interesting results. The sample of pPN
candidates was found to contain about a dozen objects overabundant in
heavy metals synthesized via neutronization of iron nuclei under
conditions of low neutron density ($s$-process). The physical conditions
necessary for efficient $s$-process and subsequent dredge-up of the matter
enriched in heavy nuclei can be found just in stars at the AGB stage (for
the history and modern understanding of the problem see the review by
Busso~et~al. [\cite{Busso}]). An analysis of the spectra of the sample of
pPN objects showed that the expected overabundances of $s$-process
elements are observed in the atmospheres of those pPN stars whose
atmospheres are carbon enriched and whose IR spectra exhibit an emission
at 21\,$\mu\mbox{m}$ [\cite{Klochkova1995,
Klochkova1998,Winckel,IRAS20000}]. The overwhelming majority of pPN stars
exhibit neither carbon (O-rich stars), nor heavy-metal overabundance (see,
e.g., [\cite{Klochkova1995,IRAS19475,IRAS20056}]). The correlation found
between the overabundance of heavy metals in the star atmosphere and the
peculiarities of the IR spectrum of the star envelope should be explained,
hence the sample of well-studied pPNe should be extended.

We know little about the properties of the star V2324\,Cyg. Arkhipova
~et~al. [\cite{Arkh1994,Arkh2}] performed long-term UBV observations and
found V2324\,Cyg to be variable with an amplitude of about
$0\lefteqn{.}^{\rm m}3$ in the U band and about $0\lefteqn{.}^{\rm m}2$ in
the V and B bands. The conclusion made by Arkhipova~et~al. [\cite{Arkh2}]
about the absence of well-defined pulsation periodicity in the star's light
variations is of great importance. The above authors explain the
photometric variability of the star by the effect of the stellar wind. The
lack of pulsations was found to be consistent with the rather early
spectral type of the star -- A3\,I~[\cite{Arkh2}]. Note that SIMBAD lists a
different spectral type -- Sp\,=\,Fe. In the polarimetric
survey~[\cite{Gledhill}] the source IRAS\,20572+4919 is classified as an
object without polarization, and this fact may be indicative of a
spherical shape of the circumstellar shell. Hrivnak~et~al.~[\cite{Hrivnak}]
analyzed near-IR spectra of a sample of pPN objects and pointed out the
anomalous (flat) form of the IR spectrum and the lack of Brackett hydrogen
lines in the spectrum of IRAS\,20572+4919. Kelly and
Hrivnak~[\cite{Hrivnak2}], in their near-IR study of the pPN sample in the
2.1--2.3\,$\mu\mbox{m}$ wavelength interval, classified the central star
of the source as an F-type star and put it among the small group of stars
with a Br$\gamma$ emission. Garcia-Lario~et al.~[\cite{Garcia-1997}], who
used IR photometry to determine the evolutionary status of 225\, IRAS
sources, consider the classification of the IRAS\,20572+4919 source as a
``post-AGB'' star to be tentative.

So far only low-resolution optical spectroscopy was performed for
V2324\,Cyg, however, even low-resolution observations revealed the main
peculiarities of the optical spectrum of V2324\,Cyg -- NaI\,D-line
emission and especially powerful emission in H$\alpha$. The H$\alpha$
emission was found more than 30 years ago~[\cite{Wishnew}]. Emission is
naturally associated with the presence of a circumstellar shell, which
affects H$\alpha$ and NaI\,D-line profiles in the optical spectrum.
Recently, a 2\AA{}-resolution spectrum of V2324\,Cyg has been
published~[\cite{Pereira}], which shows, in particular, the presence of
emission in the lines of the OI\,$\lambda$\,7773\,\AA{} triplet.

In this paper we present the results of a spectroscopic monitoring
of  V2324\,Cyg with high spectral resolution performed with the
6-m telescope of the Special Astrophysical Observatory of the
Russian Academy of Sciences (SAO RAS). The aim of our study is to detect
the spectroscopic variability, analyze the velocity
field in the atmosphere and envelope of the star, determine the
fundamental parameters, metallicity and chemical composition, and
refine the evolutionary status of the star.
In Section\,\ref{observ} we briefly describes the methods
of observation and data reduction; in Section
\,\ref{results} we reports and analyzes the results
obtained, and finally in Section\,\ref{conclus} we briefly
formulates the main conclusions.

\section{Observations and data reduction}\label{obs}

Our observations of V2324\,Cyg were performed with the 6-m telescope of
the SAO RAS. The first spectrum was taken in 1995 with a spectroscopic
resolution of R=25000 with the Lynx---echelle spectrograph~[\cite{lynx}]
mounted in the Nasmyth focus. PFES combined with a 1K$\times$1K CCD,
recorded the wavelength interval $\lambda\lambda$ 4720--6860\,\AA. The
second spectrum was taken with the Primary Focus Echelle Spectrograph
PFES~[\cite{pfes}] (R\,=\,15000). This spectrograph, combined with a
1K$\times$1K CCD, recorded the wavelength interval $\lambda\lambda$
4680--8590\,\AA. All the subsequent observations of the star we made 
in the Nasmyth focus with NES echelle spectrograph~[\cite{nes}],
which provides a spectral resolution of R=60000. Observations were
performed using a large-format 2048$\times$2048~CCD and with an image
slicer~[\cite{nes}]. Table~\ref{observ} lists the folowing data for each
spectrum: date, average UT observing time, integration time,
wavelength interval, and the spectrograph employed.

We extracted the data from the two-dimensional echelle spectra using the
modified ~[\cite{Yushkin}] context of MIDAS package. Cosmic-ray hits were
removed via median averaging of two spectra taken consecutively one after
another. Wavelength calibration was performed using the spectra of a
hollow-cathode Th-Ar lamp. To perform subsequent reduction including
spectrophotometric and position measurements we used DECH20~[\cite{gala}]
program. 

\section{Discussion of results}\label{results}

\subsection{Peculiarities of the optical spectrum of V2324\,Cyg}

We obtained most of our spectroscopic data described in
Table\,\ref{observ} with the limiting spectral resolution
R\,$\approx$\,0.1\,\AA{}. Even a preliminary examination of the
high-resolution spectra showed that V2324\,Cyg is an F-type star with wide
absorptions and P\,Cyg-type profiles for some lines (in the recorded
wavelength interval these are H$\alpha$, H$\beta$,  D1 and D2
NaI lines). Besides the expected emissions in H$\alpha$ and in the lines of
the NaD doublet, we found another spectral peculiarity -- the large widths
of absorption lines, which is untypical for a star with the post-AGB
status.

{\it Spectral type.} Arkhipova and Ikonnikova~[\cite{Arkh1994}] obtained
an indirect estimate of the star's spectral type, B8--A2\,Ib--II, based on
the results of photometric observations. Pereira and
Miranda~[\cite{Pereira}] found a later spectral type from a low-resolution
spectrum, but report equally high luminosity. The above authors chose the
HD\,842 star (an A9\,I-type supergiant) from the atlas of Jacoby~et~al.
~[\cite{Jacoby}] to be closest to V2324\,Cyg in terms of spectrum.
However, the classification of V2324\,Cyg given by Pereira and
Miranda~[\cite{Pereira}] is not satisfactory for two reasons:
\begin{enumerate}

\item{The spectrum of V2324\,Cyg reported by Pereira and
Miranda~[\cite{Pereira}] is also very similar to another spectrum of atlas
of Jacoby~et~al. ~[\cite{Jacoby}] -- that of HD\,64191 F0-3\,III (the
$\delta$\,Sct-type variable AD\,CMi). The depth ratio of the G-band to
H$\gamma$ allows V2324\,Cyg to be classified even as an F5-type star,
however, below we show that Balmer lines in the spectrum of V2324\,Cyg are
anomalous.}

\item{The result of classification may be distorted by the widths of
absorptions~[\cite{Gray}], which differ strongly for V2324\,Cyg and the
comparison stars. The rotation velocities $V\sin i$ of V2324\,Cyg,
HD\,842, and HD\,64191 are equal to 69\,km/s (as measured from our
spectra), 16\,km/s (also our estimate, but based on a spectrum adopted
from ELODIE.3 ~[\cite{Prugniel}] and ELODIE.3.1~[\cite{Prugniel2007}]
libraries), and 12\,km/s~[\cite{Solano}], respectively.}
\end{enumerate}

Our data allowed us to perform spectral classification based on line
intensity ratios of neutral to ionized atoms. We primarily used
numerous Fe\,I and some of the Fe\,II absorptions. To build the calibration
curves, we used the data from the atlas of Klochkova~et~al.~[\cite{Atlas}]
and ELODIE.3 spectral libraries~[\cite{Prugniel}]. The latter paper is
especially convenient for us because it contains spectra with the lines of
approximately the same width as those in the spectrum of V2324\,Cyg and we
can directly compare their fragments and profiles of lines and blends and
compare the line depths rather than the equivalent widths.

Within the spectral type F absorption depth ratios in the FeII/FeI line
pairs decrease from F0 to F8 and they are higher and decrease more
abruptly for supergiants than for main-sequence stars. The depth ratios
for V2324\,Cyg correspond to a spectral type interval from A9--F0\,V to
F8--G0\,I. The supergiant option must be rejected, because it would imply
too large line depths for the depth ratios in question.

The spectra of ELODIE.3 library~[\cite{Prugniel}] for stars with rotation
velocities in the 60 to 80\,km/s interval -- HD\,58923 (F0\,III),
HD\,47072 (F0\,II), HD\,432 (F2\,III(IV)), and HD\,219877 (F3\,IV) --
reproduce our spectra of V2324\,Cyg fairly well. The above stars have bona
fide parallaxes (ranging from 9 to 60\,mas according to HIPPARCOS catalog)
and direct estimates of their absolute magnitudes yield 0.2$<$\,${\rm
M_v}$\,$<$2.8$^{\rm m}$. These estimates are consistent with the MK
spectral types for the three stars and the only exception is HD\,47072
whose parallax is indicative of a III--IV luminosity class.

A comparison of the mean central depths R$_{\rm o}$ of absorption lines
averaged over the spectra of the comparison stars mentioned above with the
mean depths averaged over our spectra of V2324\,Cyg show that, with a few
strongest lines excluded, these averaged R$_{\rm o}$ agree within errors
to both for FeI and FeII. It is thus safe to assume that V2324\,Cyg is an
early F-type giant or subgiant. We estimate the spectral type more
accurately by analyzing the dependences of the R$_{\rm o}$(FeII)/R$_{\rm
o}$(FeI) ratios on spectral type Sp obtained for stars of luminosity
classes III and IV. Thus, e.g., the depth ratio for the pair of closely
located lines FeII(43)\,$\lambda$\,4732 and FeI(554)\,$\lambda$\,4737
decreases from 1.3 to 0.9, and that for the FeII(40)\,$\lambda$\,6433,
FeI(62)\,$\lambda$\,6431 blend, from 1.6 to 0.7 as one passes from F0\,III
to F5\,III. The corresponding ratios for V2324\,Cyg are equal to 1.05 and
0.98, respectively. Based only on moderate-intensity metal lines, we 
estimate the spectral type of V2324\,Cyg to be F2\,III.

\begin{figure*}[tbp]
\includegraphics[width=5.6cm,angle=-90]{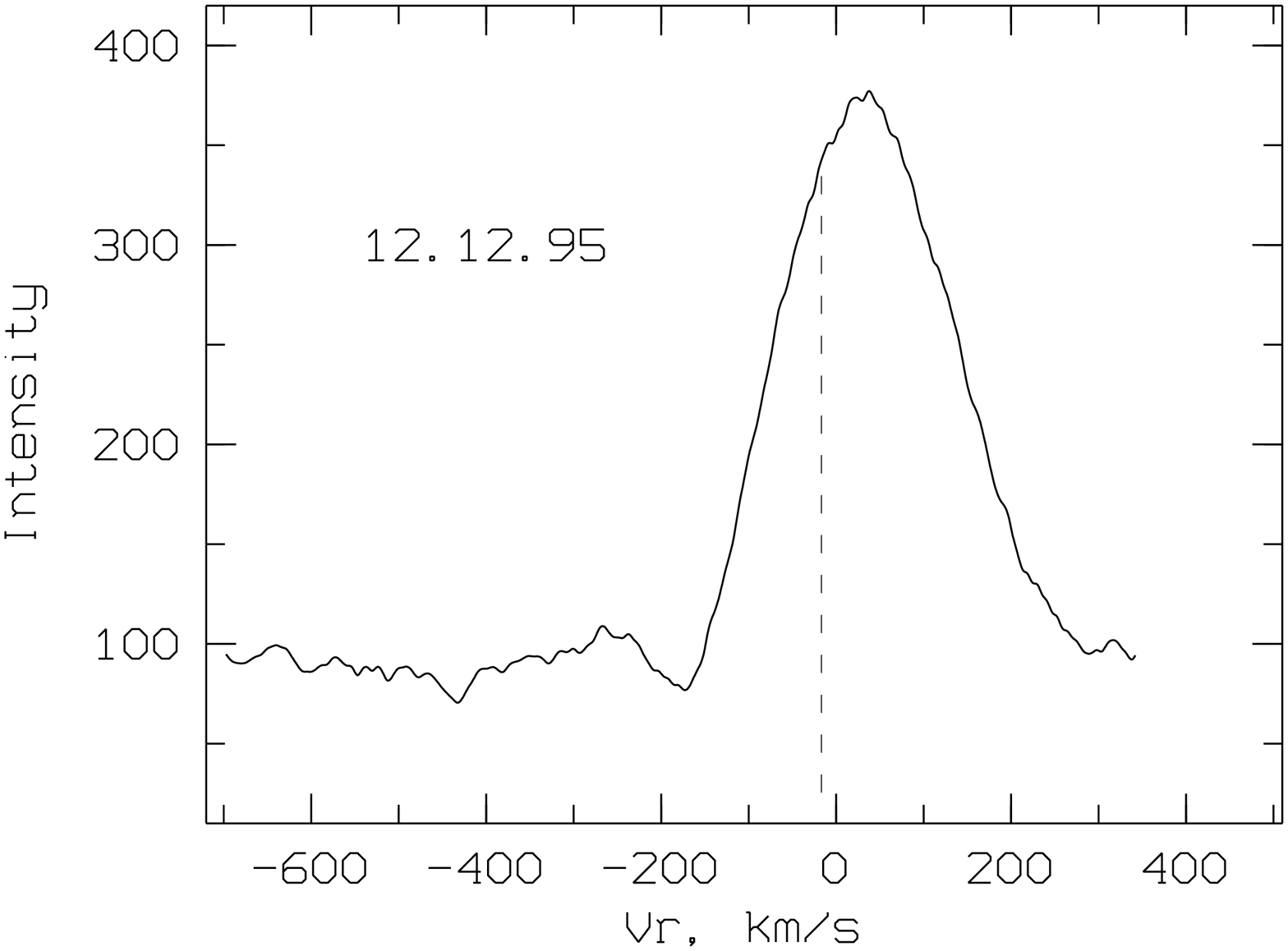}
\includegraphics[width=5.6cm,angle=-90]{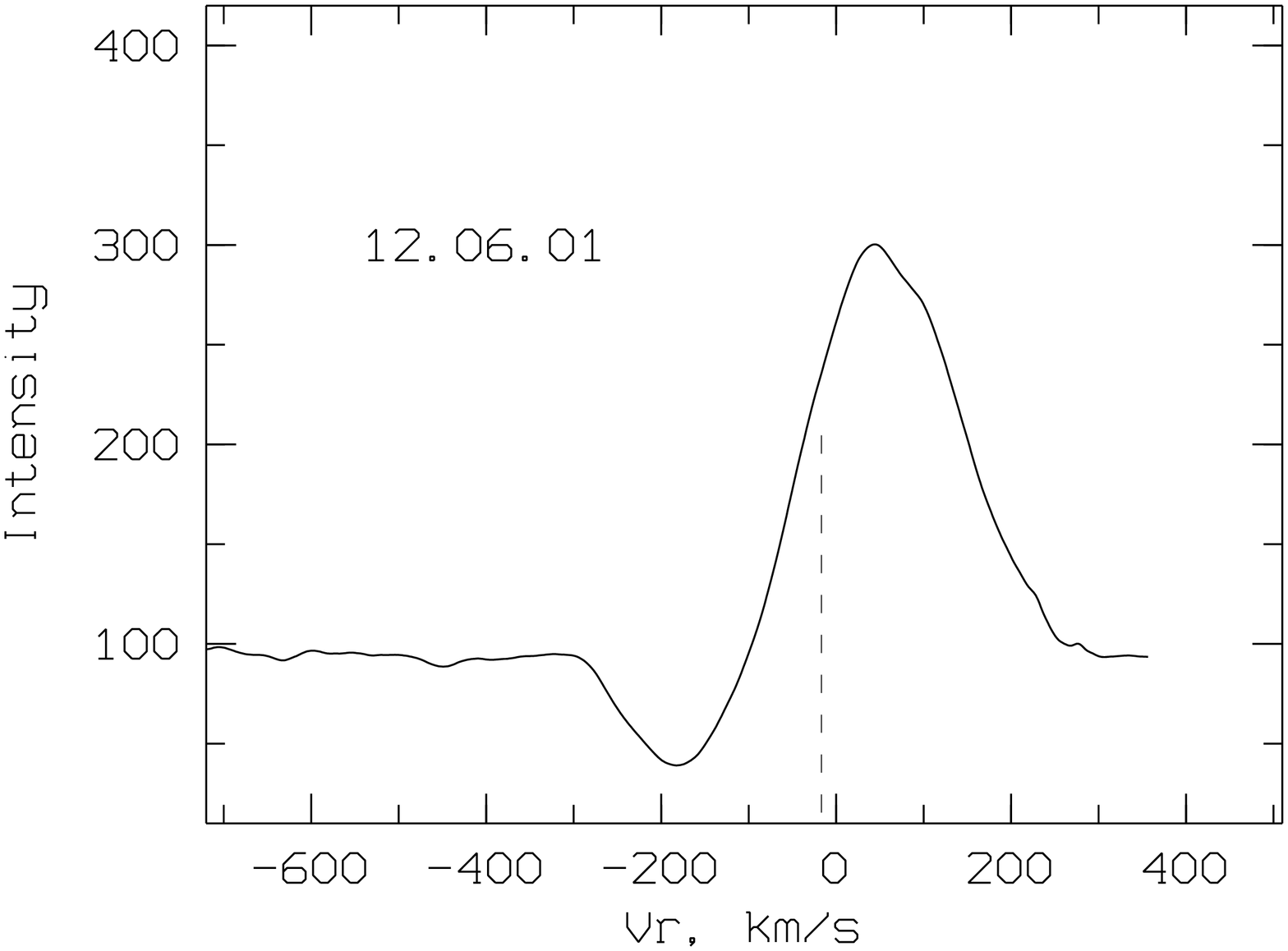}
\includegraphics[width=5.6cm,angle=-90]{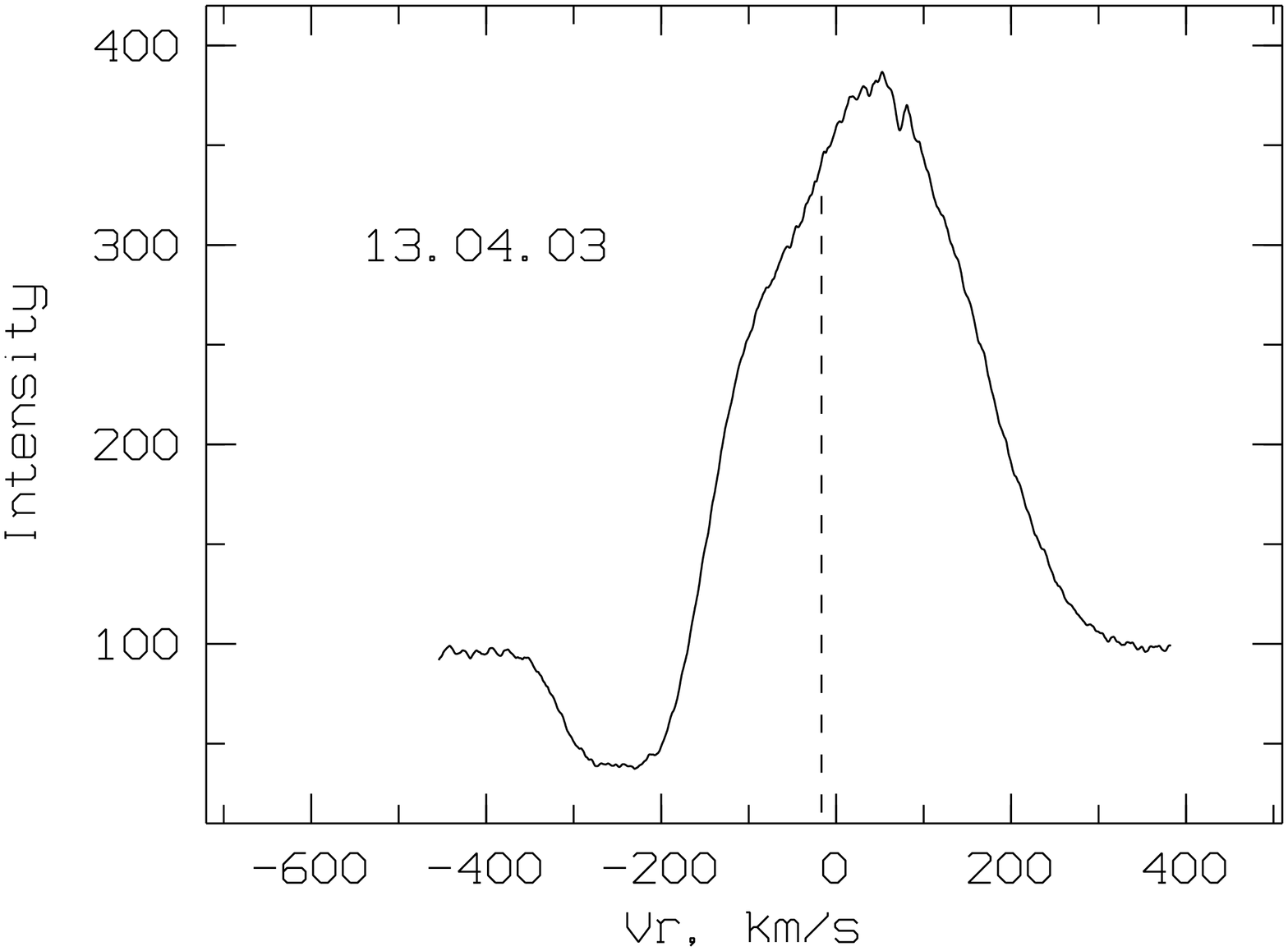}
\includegraphics[width=5.6cm,angle=-90]{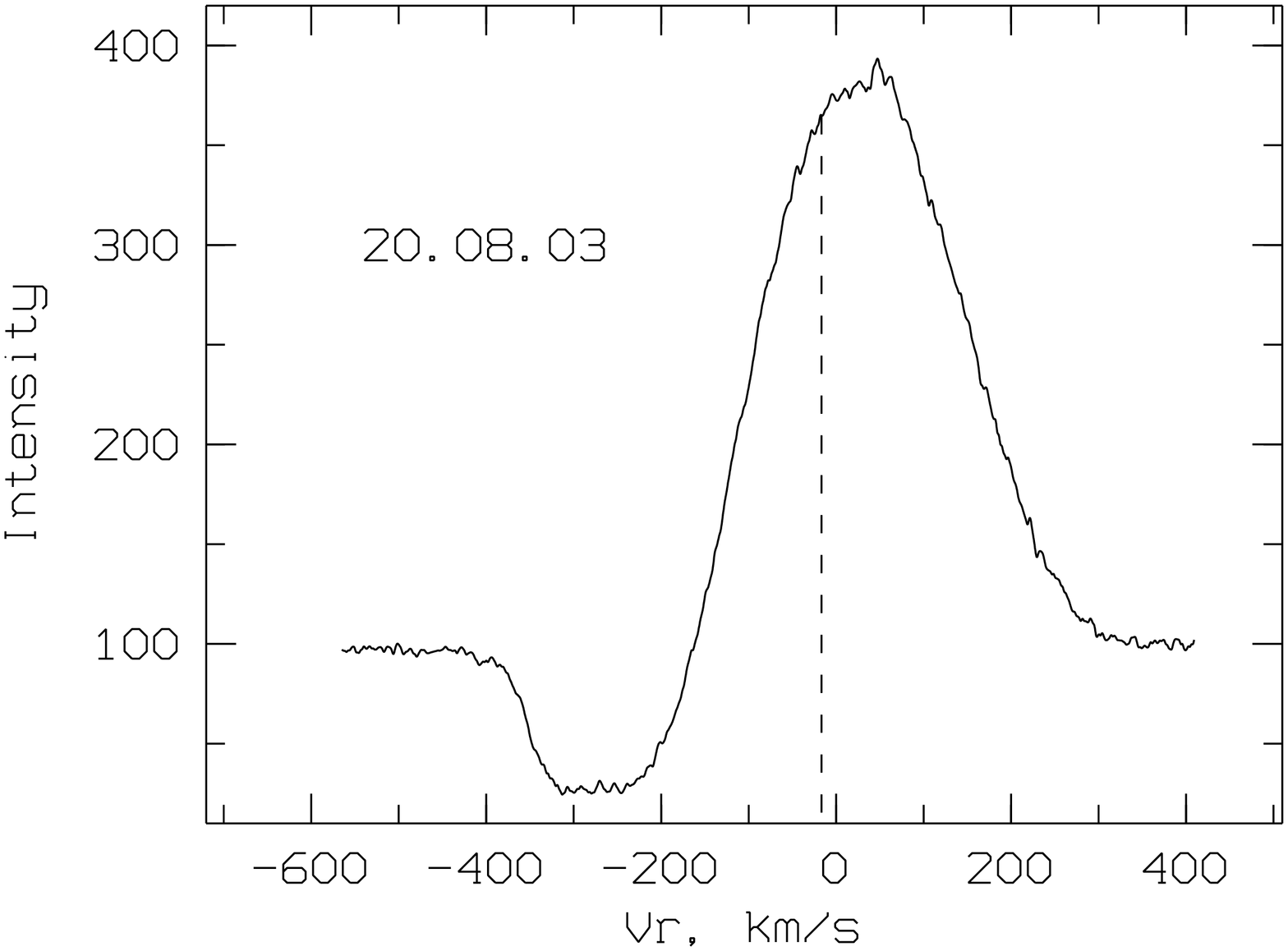}
\includegraphics[width=5.6cm,angle=-90]{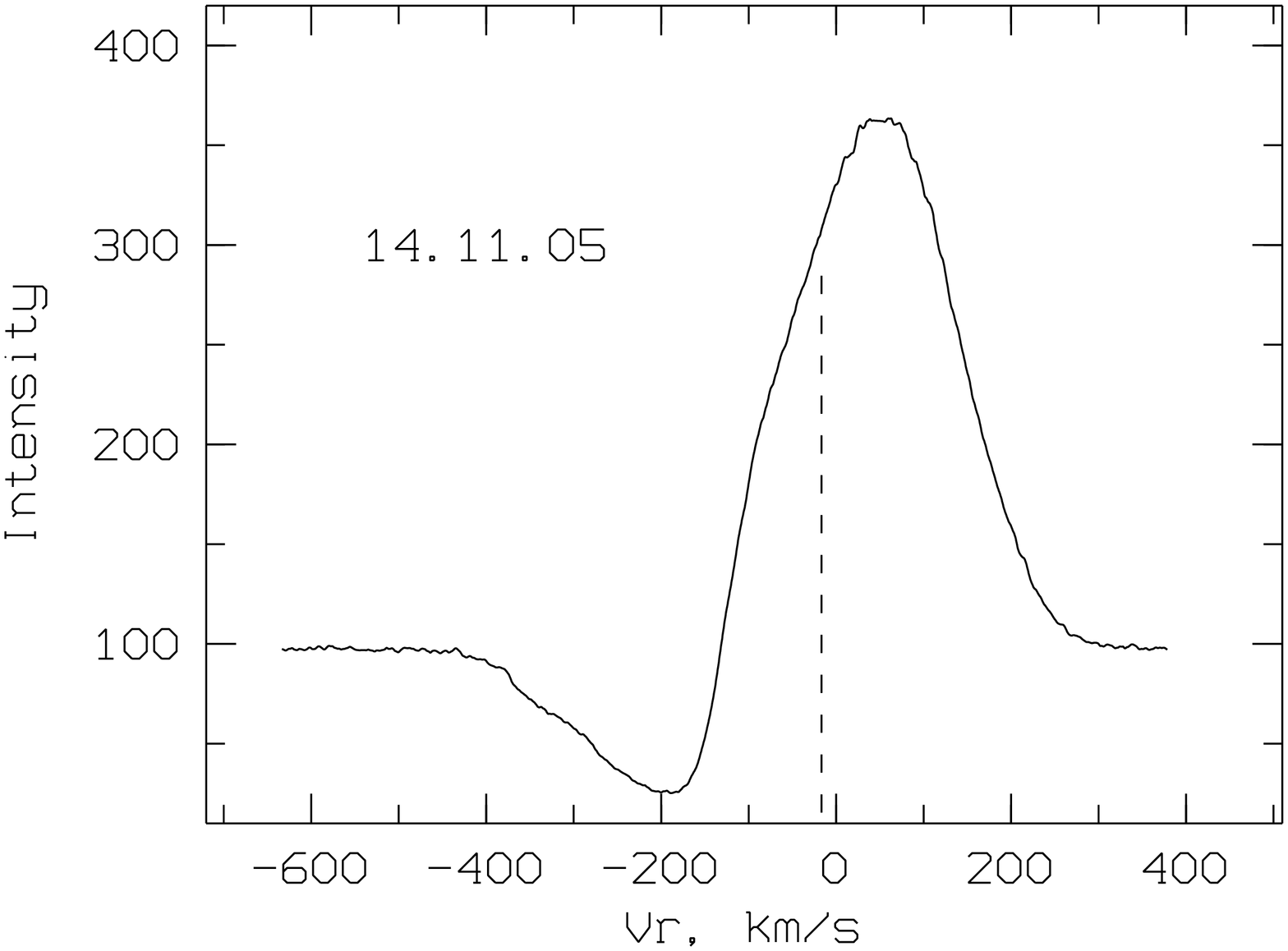}
\includegraphics[width=5.6cm,angle=-90]{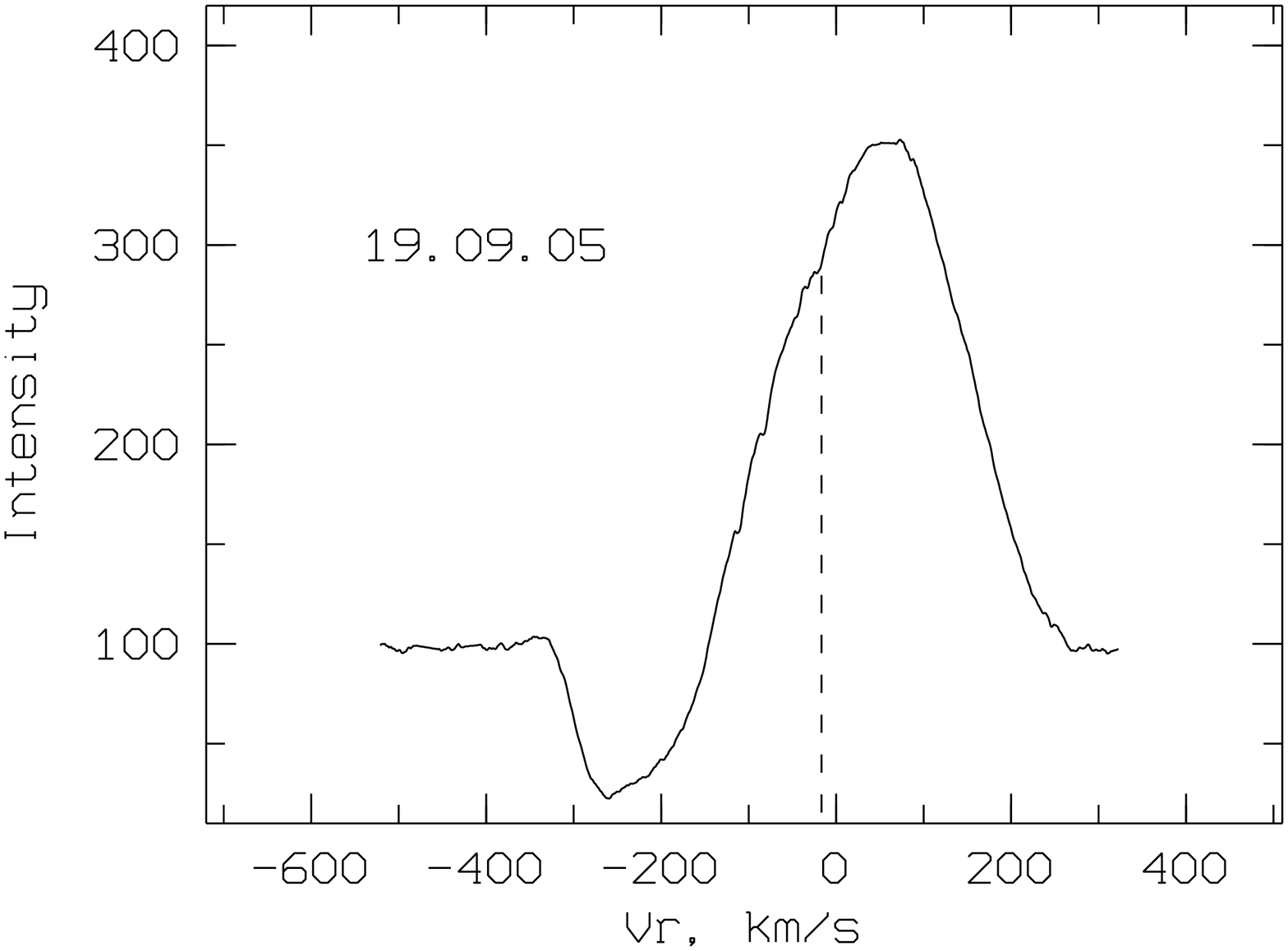}
\includegraphics[width=5.6cm,angle=-90]{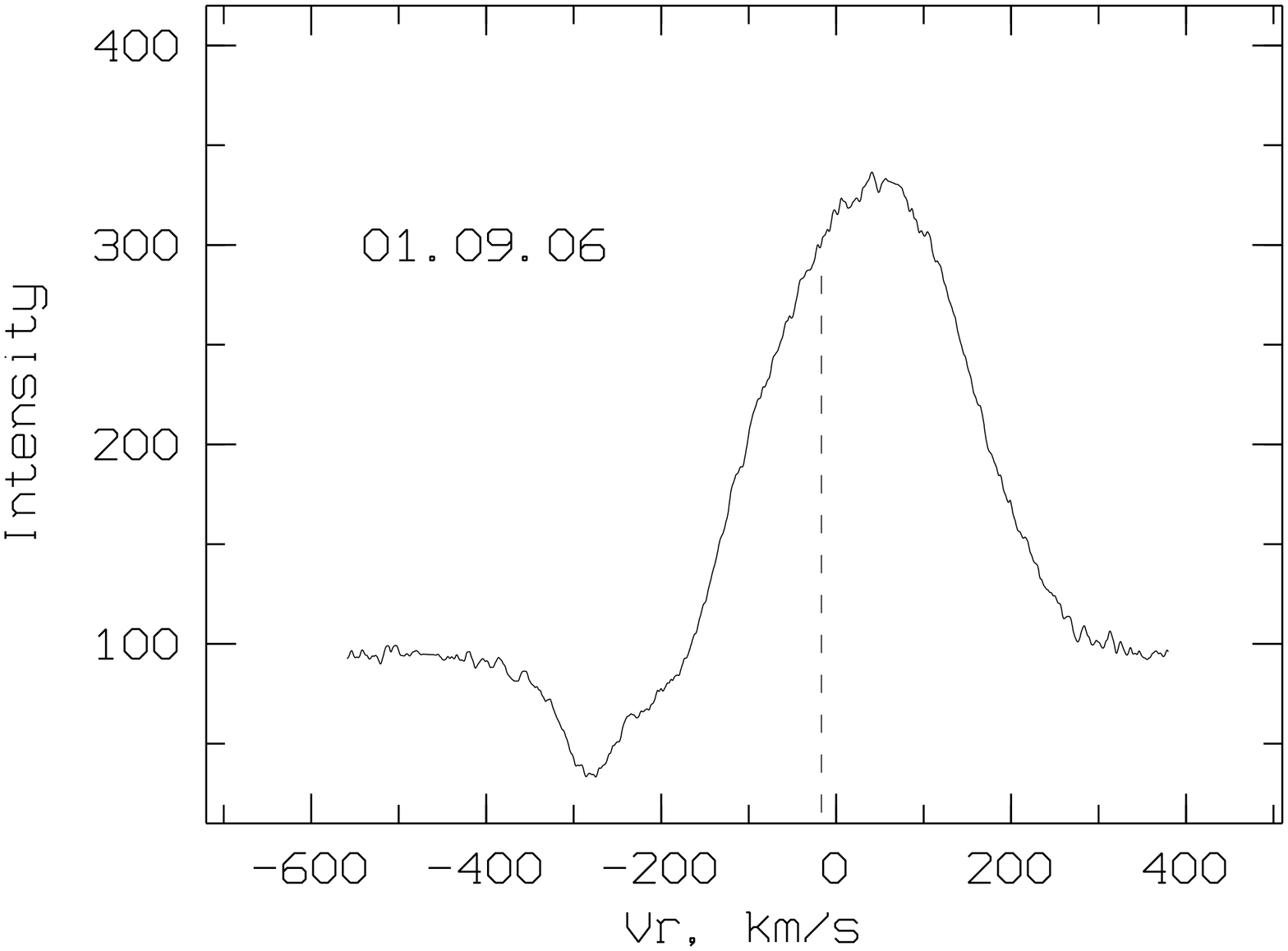}
\captionstyle{normal} \caption{H$\alpha$-profile variability in
the spectra of  V2324\,Cyg. The dashed line shows the mean
velocity obtained from photospheric lines of metals. Here and in
the following figures the continuum level corresponds to 100.}
\label{Halpha}
\end{figure*}

\begin{figure}[tbp]
\includegraphics[width=9.5cm,angle=-90]{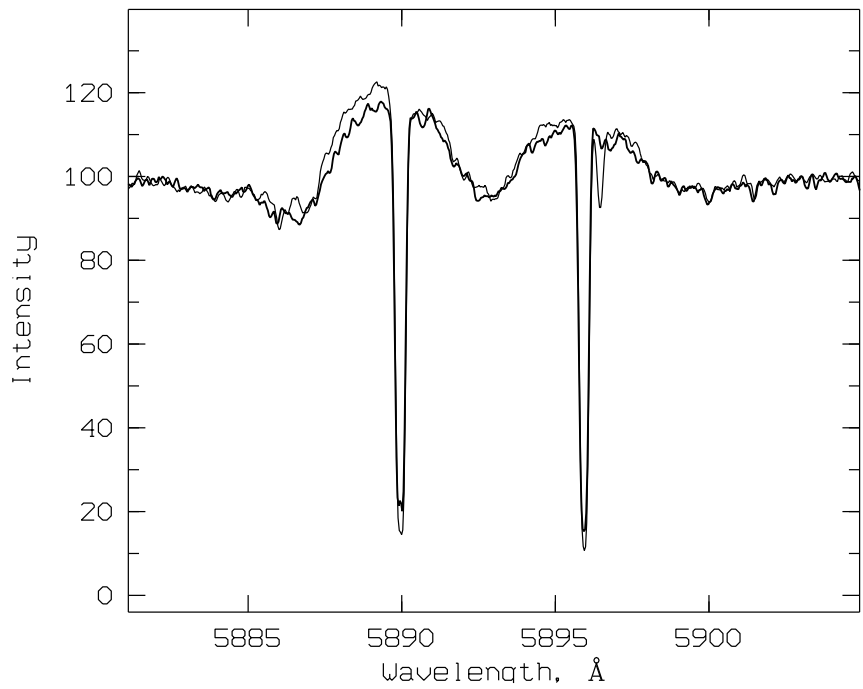}
\caption{The profile of NaI~D lines in two spectra of  V2324\,Cyg taken in 2005} \label{NaD}
\end{figure}

{\it Spectral anomalies}. Some of spectral features make V2324\,Cyg differ
markedly from a common early F-type giant. These are, first and foremost,
emissions in a number of lines. In the wavelength interval accessible for
us such emissions can be directly seen in H$\alpha$ (see
Fig.\,\ref{Halpha}) and in the NaI(1) doublet as red shifted components of
their P\,Cyg-type profiles (see Fig.\,\ref{NaD}). The peaks of emissions
in NaI lines are cut by absorptions, however, we can adopt r$\approx$1.25
as the lower limit of residual intensity for D2 (Fig.\,\ref{NaD}). It
follows from Fig.\,\ref{Halpha} that the peak of H$\alpha$ emission in our
spectra is higher than the continuum level at least by a factor of three
(and in two spectra, at least by a factor of four). A comparison of
H$\alpha$ profiles in spectra taken at different times is indicative of
their variability: both the shape and intensity of emission vary from one
spectrum to another and so do the positions of the emission and absorption
peaks.

The H$\beta$ also has a weak emission component whose intensity and
position vary with time (see Fig.\,\ref{Hbeta}). The presence
of emission shows up in the asymmetry of the profile. If we cut the
profile into two parts by the vertical line ${\rm V_r}$\,=\,0, then its
red part is raised and the blue part, on the contrary, depressed. At ${\rm
V_r}$$\approx$100\,km/s the depth of the profile R$\approx
\mbox{0.1--0.2}$, and on August 20, 2003 it even reached the continuum level.
At the same time, the blue shifted core of absorption at ${\rm
V_r}$$\approx -(\mbox{100--200}$)\,km/s is deeper than the
corresponding feature in the spectra of any comparison star
mentioned above (June 12, 2001 R$_{\rm o}\approx$0.85).

The OI(1)\,$\lambda$\,7774 blend in the low-resolution spectrum of Pereira
and Miranda~[\cite{Pereira}] appears as emission, which is at least
comparable to NaI(1) doublet in terms of intensity. In our only spectrum
taken on January 12, 2001 the OI(1)\,$\lambda$\,7774 blend also appears to
be partly filled by emission: the equivalent width of the blend is
W$\approx$0.4\,\AA{}, which, according to
Faraggiana~et~al.~[\cite{Faraggiana}] corresponds to luminosity class V for
early F-subtypes and is twice lower than for an average F2\,III-type star.

Moreover, as we pointed out above, for comparing the depths of metal
absorption lines in the spectra of V2324\,Cyg and those of comparison
stars we discarded the strongest lines -- first and foremost the FeII(42)
and MgI(2) triplets. The average central depths $R_{\rm o}$ for the
comparison stars are~$\approx 0.3$, whereas for V2324\,Cyg R$_{\rm
o}$(FeII)$\approx 0.12$ and R$_{\rm o}$(MgI)$\approx 0.16$. Almost all
these lines are parts of close blends and dominate in these blends in the
spectra of normal stars. However, in the spectrum of V2324\,Cyg these
lines are weakened and reduced to the level of other contributors (mostly
FeI). We may try to attribute these weaker intensities to the presence of
emissions, however, this hypothesis fails to explain why reduced intensity
of the lines considered does not result in appreciable blueshifts of
FeII(42) and MgI(2) absorptions relative to weak and intermediate metal
lines (like in the case of the OI(1) triplet and in contrast to H$\beta$).

\subsection{Radial velocities}

The velocities that we discuss below are measured only from minimally
blended lines. Table~2 lists the main and in some cases also the blending
components for these lines, their central residual intensities r, and
heliocentric radial velocities ${\rm V_r}$.

We found no significant date-to-date variations of the depths and
velocities for selected weak and intermediate (photospheric) absorptions.
We cannot rule out low-amplitude velocity variations (at a level of
1--2\,km/s), but these variations may also be due to pulsations, binarity,
or unaccounted systematic measurement errors. Therefore in Table\,2 we
list for these stars only the mean ${\rm V_r}$ values averaged over the
entire observational data. For the H$\beta$, NaI(1), and H$\alpha$ lines,
which exhibited profile variability, Table\,2 gives the r and ${\rm V_r}$
values for each observing date. For the latter two lines the ``/'' sign
separates the quantities corresponding to the absorption and emission
components of their P\,Cyg-type profiles.

The depths of selected photospheric absorptions in Table\,2 lie in the
interval 0.20$<$R$_{\rm o}<$0.03. Radial velocity ${\rm V_r}$ does not
exhibit appreciable systematic variation with line depth R${\rm _o}$,
thereby justifying the computation of the mean velocity ${\rm V_r}$=$-16.8
\pm 0.6$\,km/s. This velocity must be close to the systemic velocity
of the star, $\rm V_{sys}$. No reliable determination of systemic velocity
V$_{\rm sys}$ is available for V2324\,Cyg, because no CO molecular-band
~[\cite{Hrivnak3}] or H$_2$O maser~[\cite{Suarez}] emission have been
detected from the star.

Velocities inferred from the cores of the absorption components of the wind lines vary
with time. As is evident from Table\,2, in the spectra containing H$\beta$ and NaI\,D2  the
absorption cores of these two lines yield close velocities, which vary from  $-140$ to
$-225$\,km/s (and the expansion velocities of the corresponding layers, from about
120 to 210 km/s). The cores of  H$\alpha$ lines yield lower radial velocities: they vary from
$-180$ to $-280$\,km/s. The highest expansion velocity (in the spectral domain recorded in our
observations) is estimated from the blue boundaries of  H$\alpha$ absorptions: it was as high
as 450\,km/s for the blue component of split absorption measured on December 12, 1995.

\subsection{Parameters and chemical composition of the atmosphere}\label{model}

Determination the fundamental parameters from spectroscopic data and of
the elemental abundances are difficult to perform in case of V2324\,Cyg
because of the large width of absorptions in its spectrum caused by large
rotation velocity $V\sin i=69\,\mbox{km/s}$. Below we show that the errors
of measured equivalent widths are the main contributors to the uncertainty
of the chemical composition.

We use the grid of model hydrostatic and LTE atmospheres for different
metallicity values computed by Shulyak~et~al.~[\cite{Tsymbal}] to
determine the parameters of the star's model atmosphere -- effective
temperature and surface gravity -- needed to compute the chemical
composition and synthetic spectra. Fixing the main parameters -- effective
temperature $T_{eff}$ and surface gravity $\log g$ -- is known to be always
the most difficult part of the computation of the chemical composition of
a star. It is difficult to use the photometric data for the finding the
effective temperature for objects with uncertain evolutionary status and
hence with uncertain reddening. As we noted above, the H\,I Balmer-line
profiles in the spectrum of V2324\,Cyg are distorted by variable emission.
That is why we determined the effective temperature of the star from the
condition that FeI abundance must be independent on the excitation
potential of the corresponding lines; chose the surface gravity based on
the condition of ionization balance for iron atoms and ions. The
microturbulence velocity $ \xi_t$ was determined from the condition that
iron abundance must be independent of line intensity. Because of the
likely distortion of metal line profiles due to emission, for the above
procedures we selected only the lines, where such distortions were
visually small or lacking.

The validation criterion for the method is the lack of the dependence of
the abundance on the excitation potential of the corresponding lines for
other chemical elements represented by numerous lines (ScII, TiII, CrII).
Moreover, in the case of a bona fide values of the microturbulence
velocity individual abundances must be independent on the equivalent
widths of the lines used in the computation. The typical accuracy of model
parameters for a star with effective temperature of about 7500\,K is
$\Delta T_{eff} \approx 200$\,K, $\Delta \log g \approx 0.5\,\mbox{dex} $,
and $\Delta \xi_t \approx 1.0$ km/s. Table\,\ref{chem} lists the
abundance errors due to the parameter errors mentioned above.
From a comparison of the errors in Tables\,\ref{chem} and
\ref{error} one can see that the errors of equivalent line width
measurements are the main contributors to the uncertainties for the most
abundances.

\begin{figure}[tbp]
\includegraphics[angle=-90,width=0.75\textwidth]{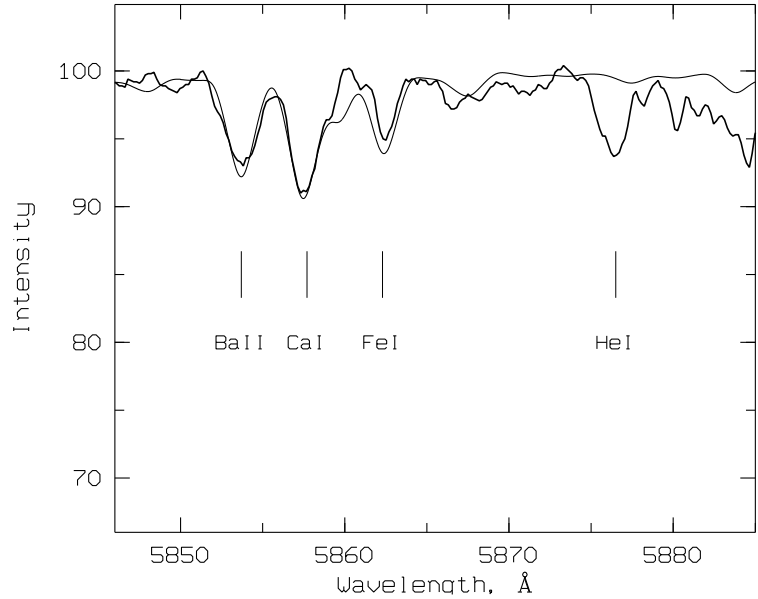} 
\caption{BaII\,$\lambda$\,5853 and HeI\,$\lambda$\,5876 lines in the spectrum of  V2324\,Cyg
        taken in 2001. The thin line represents the theoretical spectrum computed for
        $T_{eff}$\,=7500\,K, $\log g$\,=\,2.0, and $\xi_t$\,=\,6.0\,km/s and solar elemental
        abundances.}
\label{He}
\end{figure}

\begin{figure}[tbp]
\includegraphics[angle=-90,width=0.75\textwidth]{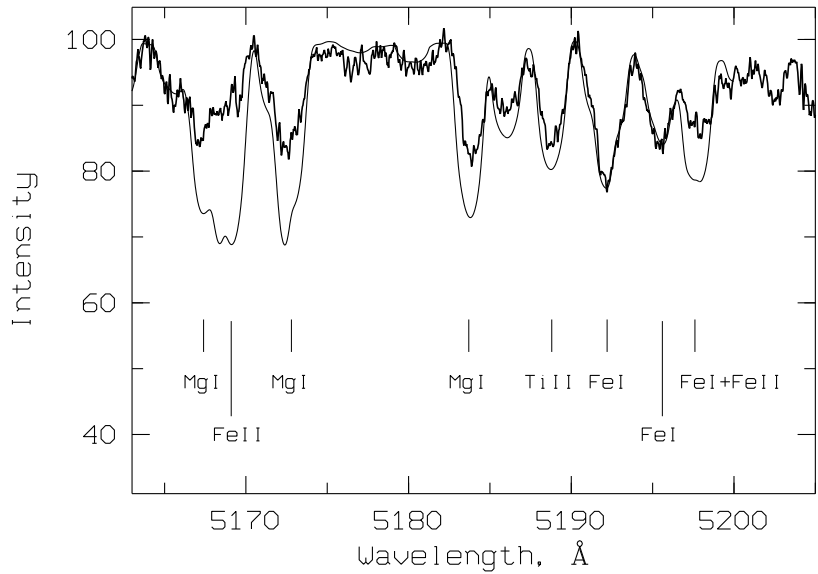} 
\caption{MgI\,$\lambda$\,5167--5183 magnesium triplet and FeII\,$\lambda$\,5169 iron
         lines in the spectrum of  V2324\,Cyg taken on November 13, 2005. The thin line represents the
         theoretical spectrum computed for $T_{eff}$\,=7500\,K, $\log g$\,=\,2.0, and
         $\xi_t$\,=\,6.0\,km/s and solar elemental abundances.}
\label{MgI}
\end{figure}

The known test for the internal consistency of the inferred parameters
consists in an agreement between the observed and synthetic spectra.
We compared the observed spectrum with the synthetic spectrum that we
computed using the code of Shulyak~et~al.~[\cite{Tsymbal}]. For 
computations we adopted the model atmosphere with solar chemical
composition and with the following parameters: $T_{eff}=7500$\,K, $\log g
=2.0$, and $\xi_t=6.0$\,km/s. A comparison of the observed and theoretical
spectra (see examples in Figs.\,\ref{He} and \ref{MgI}) demonstrates their
satisfactory agreement provided that we neglect HeI lines and the FeII(42)
and MgI(2) lines mentioned above, which are distorted by emission.

We adopted the oscillator strengths $\log gf$ and other atomic constants
required for computing the elemental abundances from VALD
database~[\cite{VALD1, VALD2}]. Table\,\ref{error} lists the adopted
parameters of the model atmosphere $T_{eff}=7500$\,K, $\log g =2.0$,
$\xi_t =6.0$\,km/s. We used the solar chemical composition, with respect
to which we compute the elemental abundances of the star under study, from
Asplund~et~al.~[\cite{Grev}]. We performed all our chemical composition
computations using the software developed by Shulyak~et~al.
~[\cite{Tsymbal}] and adapted by the same authors to PCs operating under
OS Linux. We computed plane-parallel LTE models using the software described
by Shulyak~et~al.~[\cite{Tsymbal}] without hyperfine structure and
isotopic-shift corrections, which increase the widths of NiI and BaII
lines.

The scatter of the elemental abundances inferred from the set of spectral
lines is rather small: the standard deviation ${\sigma}$ does not exceed
0.3\,dex (see Table\,\ref{error}). We determined the parameters of the
model atmosphere using only low and intermediate intensity lines with
equivalent widths ${W \le 0.25}$\,\AA{}, because the approximation of a
state plane-parallel atmosphere may inadequately describe the strongest
spectral features. Moreover, some of the strong absorptions in the
spectrum could be distorted by the effect of the circumstellar shell. In
the case of insufficient resolution the intensities of the shell
components add up to the intensities of the components that form in the
atmosphere. We computed the elemental abundances for an extended set of
spectral lines -- for a number of elements (SiII, TiII, FeI, FeII, BaII) we
used lines with equivalent widths exceeding the equivalent-line width
limit mentioned above.
Let us now analyze the chemical abundances by grouping elements according
to the type of their synthesis.

\begin{figure*}[tbp]
\includegraphics[angle=-90,width=0.75\textwidth]{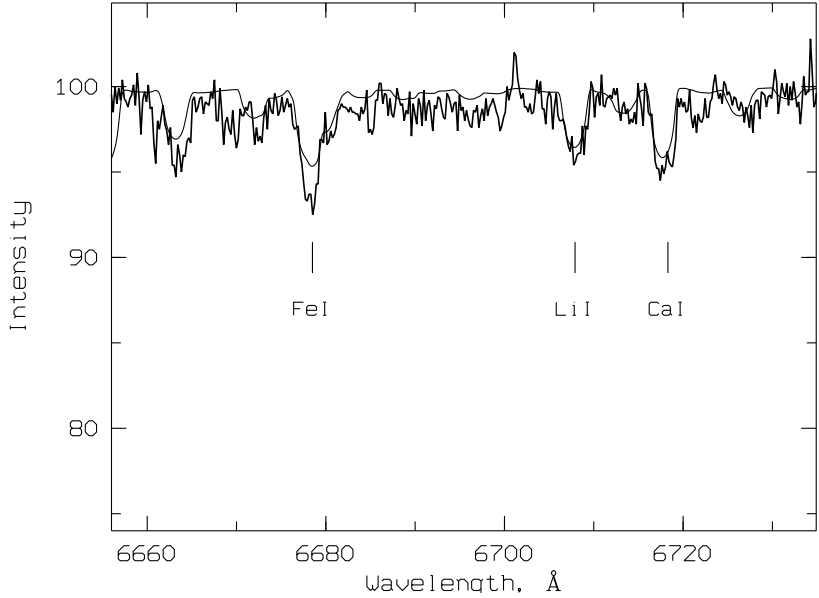}
\caption{The LiI\,$\lambda$\,6707\,\AA{} line in the spectrum of V2324\,Cyg taken on June 12,
         2001. The thin line shows the theoretical spectrum computed for $T_{eff}$\,=7500\,K,
	 $\log g$\,=\,2.0, and $\xi_t$\,=\,6.0\,km/s and the elemental abundances from
         Table\,\ref{error}.}
\label{Li}
\end{figure*}

{\it Light elements.} As one might expect in the case of an evolved star,
the abundances of a number of light elements have changed. The
overabundance of lithium computed from the LiI line $\lambda$\,6707\,\AA{}
is immediately apparent. The equivalent width of this line was measured
with confidence in all our spectra of V2324\,Cyg and its averaged
equivalent width is W$\ge$50\,m\AA{}. We therefore consider the lithium
overabundance [LiI/Fe]$\ge$+0.68 to be established beyond doubt.
Figure~\ref{Li} compares the observed and synthetic spectra of the star in
the region of the LiI\,6707\,\AA{} line. We performed this computation
with the model-atmosphere parameters $T_{eff}$=7500\,$\mbox{K}$, $\log
g$=2.0, and $\xi_t$=6.0\,$\mbox{km/s}$ and the elemental abundances
listed in Table\,\ref{error}. This is the most interesting anomaly in the
chemical abundances pattern in the atmosphere of V2324\,Cyg.

We find a small underabundance of oxygen [O/Fe]=$-0.12$ combined with a
small overabundance of carbon [C/Fe]=+0.35, implying an abundance ratio
C/O$>1$. Unfortunately, the OI\,$\lambda$\,5330.7, 6155.9, and
6158.1\,\AA{} oxygen lines employed are too weak and have their equivalent
widths determined with large errors because of the large widths of these
lines (in Table\,\ref{error} the standard error for oxygen is
$\sigma=0.28$). Therefore the inferred oxygen abundance is uncertain. No
nitrogen lines were available in the wavelength interval covered by our
observations and nitrogen abundance is of fundamental importance for the
determination of the evolutionary stage of the star. On the whole,
concerning the CNO triad the only certain result is the overabundance of
carbon.

The sodium abundance was determined from the moderate intensity NaI
$\lambda$ 5682, 5688, and 6160\,\AA~lines with small NLTE corrections
~[\cite{Takeda1,Takeda2}]. Hence the sodium overabundance found,
[Na/Fe]=+1.04, may be mostly due to the sodium synthesis in the
NeNa-cycle, which proceeds simultaneously with hydrogen burning in the CNO
cycle. The magnesium, whose abundance we determined from the two lines,
MgI\,$\lambda$\,4702.99 and 5528.40\,\AA{}, is also overabundant:
[Mg/Fe]=+0.43. The sodium-to-magnesium abundance ratio is [Na/Mg]=+0.61.

{\it Iron-peak elements.} The iron abundance, which is usually employed as
the metallicity indicator, does not differ in the atmosphere of V2324\,Cyg
from the corresponding solar abundance: $\log\varepsilon({\rm
FeI,FeII})$=7.44. The reliably enough determined abundances of chro\-mium
and nickel, which belong to the iron group, also differ only slightly from
the normal values: [CrI,CrII, NiI/Fe]=$-0.19$. The solar metallicity of
the star is consistent with the systemic heliocentric velocity adopted in
Section\,2.2, $V_{\odot}\approx -$17\,km/s, which is typical of Galaxy
disk stars.

{\it Heavy metals.} The overabundance of barium with respect to iron,
[Ba/Fe]=+0.46, is something expected, but rarely found in the atmospheres
of post-AGB supergiants. It is much more common for atmospheres of these
stars to exhibit overabundance of $s$--process
elements~[\cite{Klochkova1998,Winck}]. However, we found no overabundance of
yttrium, which is also synthesized via slow neutronization. The
overabundance of $s$-process elements or lack of such overabundance depend
on the initial mass of the star and on the mass-loss rate at the AGB
stage,which determine the evolution of the particular star and the mass of
the stellar core. Modeling of the third dredge-up process
~[\cite{HerwigAustin}] shows that the efficiency of the dredge-up of
reaction products increases with the mass of the core (and hence with the
initial mas) of the post-AGB star. It also follows from Herwig's
computations~[\cite{Herwig2000}] that the efficiency of the dredge-up
increases if we take into account penetrative convection
 at the base of the convective zone.

{\it Separation of chemical elements in the shell.} As is well known,
selective separation of chemical elements in stars with gaseous-dusty
shells may be an efficient mechanism producing anomalous elemental
abundances. However, the solar iron abundance for V2324\,Cyg indicates
that no condensation onto dust grains occurs in this shell, since iron is
an element that condenses efficiently onto dust grains. Further evidence
is provided by the normal zinc abundance,  [Zn/Fe]=+0.09, which is
consistent with the typical zinc abundance [Zn/Fe]=+0.04 over a wide range
of metallicities~[\cite{Sneden}]. The lack of manifestations of selective
separation is difficult to explain for a star with IR excess due to the
circumstellar shell.

{\it Related objects}. Earlier studied objects include pPN candidates with
a chemical composition pattern similar to that found for V2324\,Cyg.
Arellano-Ferro~et~al.~[\cite{Ferro}] analyzed a sample of
post-AGB stars and found one of them---the hot star HD\,172481
(IRAS\,18384$-$2800)---also to be overabundant in lithium. The main
parameters of this star---$T_{eff}$ = 7250\,K and $\log g$\,=\,1.5---are
also similar to those of V2324\,Cyg. However, the atmosphere of
V2324\,Cyg, unlike that of HD\,172481, is overabundant with heavy metals.
Moreover, Reyniers and Van Win\-ckel~[\cite{Reyniers}]
suspected that HD\,172481, unlike V2324\,Cyg, is a binary system and hence
the two stars cannot be considered to be full analogs.

All parameters considered, the F-type star HD\,331319 (the optical
component of the IRAS\,19475+3119 source) is a closer analog of
V2324\,Cyg. This F-type star has close to solar metallicity, it is
overabundant in helium, oxygen, and light metals (Na, Si), while being
underabundant in titan and barium~[\cite{IRAS19475}]. Like in the case of
V2324\,Cyg, the atmosphere of HD\,331319, with $T_{eff}$\,=7200\,K and
$\log g$\,=\,0.5, is overabundant in lithium [Li/Fe]=+0.62. It is
interesting that Klochkova~et~al.~[\cite{IRAS19475}] detected a
HeI\,$\lambda$\,5876\,\AA\,line with the equivalent width of $W_{\lambda}
\approx 38$\,m\AA\,in the spectrum of the F-supergiant HD\,331319 and
concluded that helium in this star is of photospheric origin and hence the
products of its synthesis have been dredged up during the evolution of the
star. As for V2324\,Cyg, we cannot rule out that the
HeI\,$\lambda$\,5876\,\AA\,line may also be present in its spectrum, at
least at some observations. As one can see from Figure\,\ref{He}, where we
compare the HeI\,$\lambda$\,5876\,\AA{} line in the spectrum of V2324\,Cyg
taken on June 12, 2001 with the synthetic spectrum computed with parameters
$T_{eff}$=7500\,K, $\log g$=2.0, $\xi_t$=6.0\,km/s, the above situation is
possible. Neither of these two stars exhibits manifestations of selective
separation of chemical elements ~[\cite{Ferro,IRAS19475}].

However, despite their close effective temperatures, metallicities, and
details of chemical composition, V2324\,Cyg and HD\,331319 differ
substantially in luminosity and rotation velocity. Metal lines in the
spectrum of HD\,331319 are narrow and the star is of luminosity class Ib,
whereas lines in the spectrum of V2324\,Cyg are broadened by rotation and
it is most likely of luminosity class III. The H$\alpha$ profiles in the
spectra of these two stars also differ significantly. The H$\alpha$
profile in the spectrum of HD\,331319 is typical of post-AGB stars (see
Fig.\,1 in paper~[\cite{IRAS19475}]), whereas the P\,Cyg-type profile of the
H$\alpha$ line in the spectrum of V2324\,Cyg is closer to the
corresponding profiles in the spectra of supergiant stars. Recall that
V2324\,Cyg has a very high mass outflow velocity -- it amounts to several
hundred km/s, which is one order of magnitude higher than the typical
expansion velocities of pPN envelopes.

There is another pPN candidate, which resembles V2324\,Cyg in terms of
spectral peculiarities and chemical composition. We mean the spectroscopic
binary BD+48$^{\rm o}$1220 which is the optical component of the IR-source
IRAS\,05040+4820. The variability of the spectrum of this star has been
recently discovered by analyzing the spectra obtained with the 6-m
telescope~[\cite{IBVS}], and parameters and chemical composition for this
star were determined later~[\cite{05040}]. Besides their close
temperatures and abundances of a number of chemical elements (solar
metallicity, overabundance of lithium and sodium), the similarity of
BD+48$^{\rm o}$1220 and V2324\,Cyg also shows up in the variability of
profiles of metal lines, which contain an emission component, and in the
strong emission in the H$\alpha$ line. The hypothesis that BD+48$^{\rm
o}$1220 and V2324\,Cyg may be at close evolutionary stages is also
corroborated by the lack of spectral features in their radio spectra.
However, these two stars have a very important difference, namely,
spectral classification gives the luminosity class Ib for BD+48$^{\rm
o}$1220. The surface gravities inferred via the method of model
atmospheres also differ: $\log g$=2.0 and 0.0 for V2324\,Cyg and
BD+48$^{\rm o}$1220, respectively.

\subsection{The problem of luminosity and evolutionary status of V2324\,Cyg}

Kinematic distance determinations in the direction of V2324\,Cyg
(l$\approx$89$\lefteqn{.}^{\rm o}4$, b$\approx$2$\lefteqn{.}^{\rm o}$4) are
unreliable: radial velocity slightly depends on distance at least up to
2\,kpc ~[\cite{Brand}]. However, color excess and equivalent widths of
NaI(1) and DIB interstellar absorptions (see Table\,\ref{DIB}) exhibit
well defined increasing with distance already within the local arm. Both
parameters are appreciably smaller for V2324\,Cyg than for the members of
the large Cyg\,OB7 association the star projects on and which, according to
Humphreys~[\cite{Humphreys}], is located at a heliocentric distance of
0.8\,kpc. A coarse estimate based on the above parameters yields for
V2324\,Cyg a distance of d=0.5--0.6\,kpc, implying a low luminosity of
M$_{\rm v} \approx$1$\lefteqn{.}^{\rm m}6$. Spectrophotometric estimation
based on our spectral type (F2III) and photometry by Arkhipova and
Ikonnikova~[\cite{Arkh1994}] (V=11$\lefteqn{.}^{\rm m}7$ and
B$-$V=1$\lefteqn{.}^{\rm m}$1) yields an even smaller distance,
d=0.35--0.4\,kpc. Such a low luminosity is also indicative of the low
initial mass of the star,which is inconsistent with the detected
overabundance of lithium. According to Straniero~et~al.~[\cite{Straniero}],
lithium overabundance may arise during the evolution of the most massive
AGB stars with initial masses ${\mathcal M}\approx7{\mathcal M}_{\odot}$.

However, lithium overabundance in the atmosphere of a low-luminosity star
suggests us an alternative evolutionary stage for V2324\,Cyg. While
analyzing possible mechanisms of lithium enrichment of the Galactic
interstellar medium, Romano~et~al.~[\cite{Romano}] focus attention on the
phenomenon of overabundance of Li in the atmospheres of F-type giants
with IR excesses. Some of these Li-rich giants, which have not yet reached
the AGB stage and are observed at the RGB, have low masses ${\mathcal
M}<2.5{\mathcal M}_{\odot}$. As a mechanism for Li production 
Romano~et al.~[\cite{Romano}] suggest the ``cool bottom process'', which is
based on the production of beryllium, its dredge-up toward the base of the
convective envelope, and subsequent decay down to lithium nuclei in the
subsurface layers of the star. Here it is appropriate to cite the recent
paper by Jasniewicz~et~al.~\cite{Jasn}, who determined $\varepsilon({\rm
Li})$ for a sample of solar metallicity G--K giants and subgiants, and
found that the highest $\varepsilon({\rm Li})$ is achieved for fast
rotating stars ($V\sin i\,\ge 30\,\mbox{km/s}$). However, as we pointed
out above, spectral features in the form of emissions prevent us from
adopting the red-giant status for the star under study.

In conclusion, we have to admit that the evolution stage of V2324\,Cyg
remains unclear despite the extensive amount of the data that we obtained.
On the IR colour--colour diagram IRAS\,20572+4919 lies in domain IV
populated by planetary and protoplanetary nebulae. According to
chronological sequences of Lewis~[\cite{Lewis}], the lack of maser
emission in the OH and H$_2$O bands~[\cite{Suarez}] suggests that the
system must be approaching the stage of planetary nebula. This fact agrees
with the conclusions of Garcia-Lario~et~al.~[\cite{Garcia-1990}] and
Arkhipova~et~al.~[\cite{Arkh2}] who consider V2324\,Cyg as a post-AGB star.
However, the post-AGB stage is inconsistent with a number of properties of
the star found as a result of our research, first and foremost, the low
luminosity of the star: spectral classification points to luminosity class
III. Here to the point to recall the paper~[\cite{Garcia-1997}], where the
classification of the IRAS\,20572+4919 source as a ``post-AGB'' star is
considered as tentative. 

The H$\alpha$ profile and very high wind velocity, which are usually
found in supergiants, are also inconsistent with the post-AGB status. We
see the prospects for further refinement of the distance, mass, and nature
of V2324\,Cyg to be found in gathering data not only about the object
itself, but also about stars seen in the direction of the Cas\,OB7
association. 

\section{Conclusions}\label{conclus}

We performed high-resolution optical spectroscopy of the V2324\,Cyg
variable star -- an optical counterpart of the IR source IRAS\,20572+4919.
We identified more than 200 absorptions within the wavelength interval
4549$\div$7880\,\AA{} (these are mostly Fe\,II, Ti\,II, Cr\,II, Y\,II,
Ba\,II, and Y\,II lines) and determined the spectral type and rotation
velocity of the star to be F0\,III and $V\sin i=69\,\mbox{km/s}$,
respectively. The main peculiarity of the spectrum of the star are complex
P\,Cyg-type profiles of the H$_\alpha$ and NaI\,D lines. 

We analyzed 9 spectra taken in different years (1995--2006) and failed to
find either systematic trend of radial velocity ${\rm V_r}$ with line depth
R$_{\rm o}$, or temporal variability of ${\rm V_r}$. This result allowed
us to adopt the mean value ${\rm V_r}$=$-16.8\pm0.6$\,km/s as the systemic
velocity, V$_{\rm sys}\approx$$-17$\,km/s. We found velocities ranging
from $-140$ to $-225$\,km/s (and the expansion velocities from about 120
to 210\,km/s for the corresponding layers) from the cores of the
absorption components of the H$\beta$ and NaI wind lines. We found the
expansion velocity to be the highest for the blue component of the
H$\alpha$ split absorption: 450\,km/s for December 12, 1995.

By the model atmospheres method we determined the effective temperature,
$T_{eff}$\,=7500\,K, surface gravity $\log g$\,=\,2.0, microturbulence
velocity $\xi_t$\,=\,6.0\,km/s, and metallicity which appear to be equal
to solar one. The main peculiarity of the chemical composition of
V2324\,Cyg is the overabundance of lithium and sodium.

The solar metallicity of the star combined with V$_{\rm sys}\approx$$-17$\,km/s
allows us to classify it as an object of the Galactic disk population.
Main conclusion of this investigation is the uncertainty of the
evolutionary status of V2324\,Cyg. Its properties are not entirely
consistent with either the post-AGB or RGB stage.

\subsection*{Acknowledgments}

{\it We are much indebted to M.V.\,Yushkin for his assistance in
observations and primary data reduction. This work was supported by the
Russian Foundation for Basic Research (project 08--02--00072\,a), the
``Extended Objects in the Universe'' fundamental research program of the
Division of Physical Sciences of the Russian Academy of Sciences, and the
``Origin and Evolution of Stars and Galaxies'' program of the Presidium of
the Russian Academy of Sciences.}


\newpage

\newpage

\clearpage
\newpage

{\hoffset=0.5cm
\begin{figure*}[tbp]
\includegraphics[width=6.0cm,angle=-90]{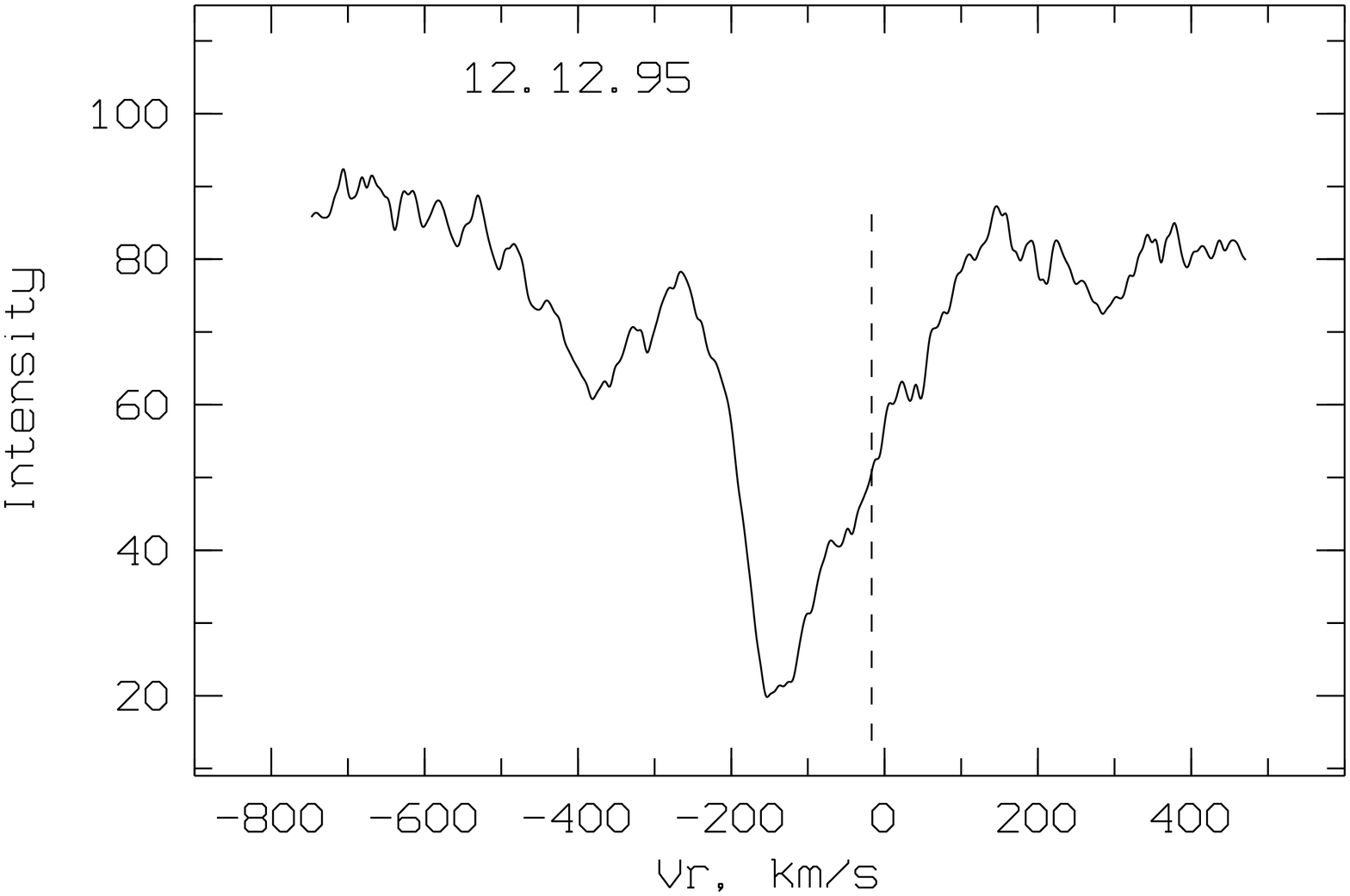}
\includegraphics[width=6.0cm,angle=-90]{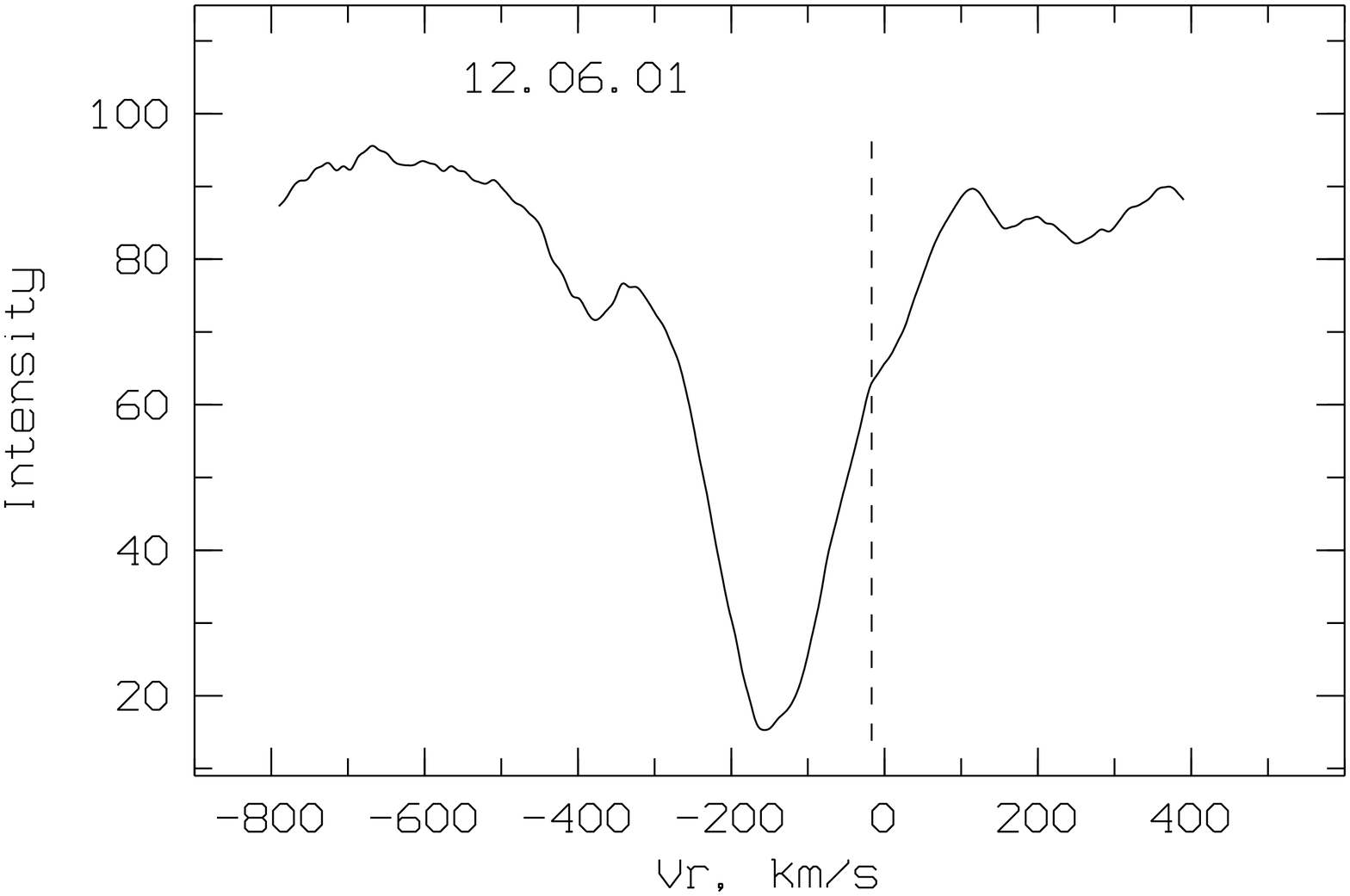}
\includegraphics[width=6.0cm,angle=-90]{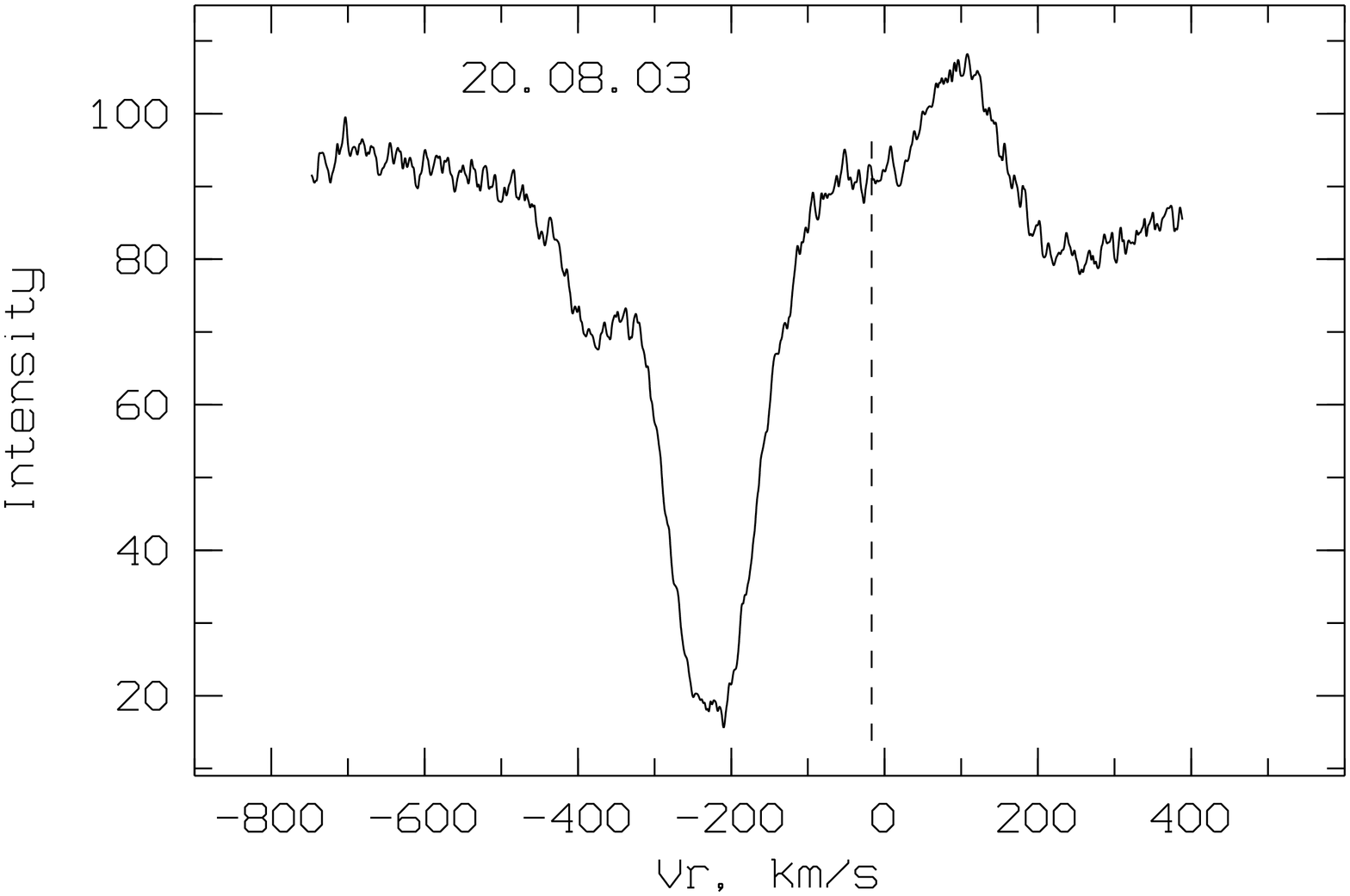}
\includegraphics[width=6.0cm,angle=-90]{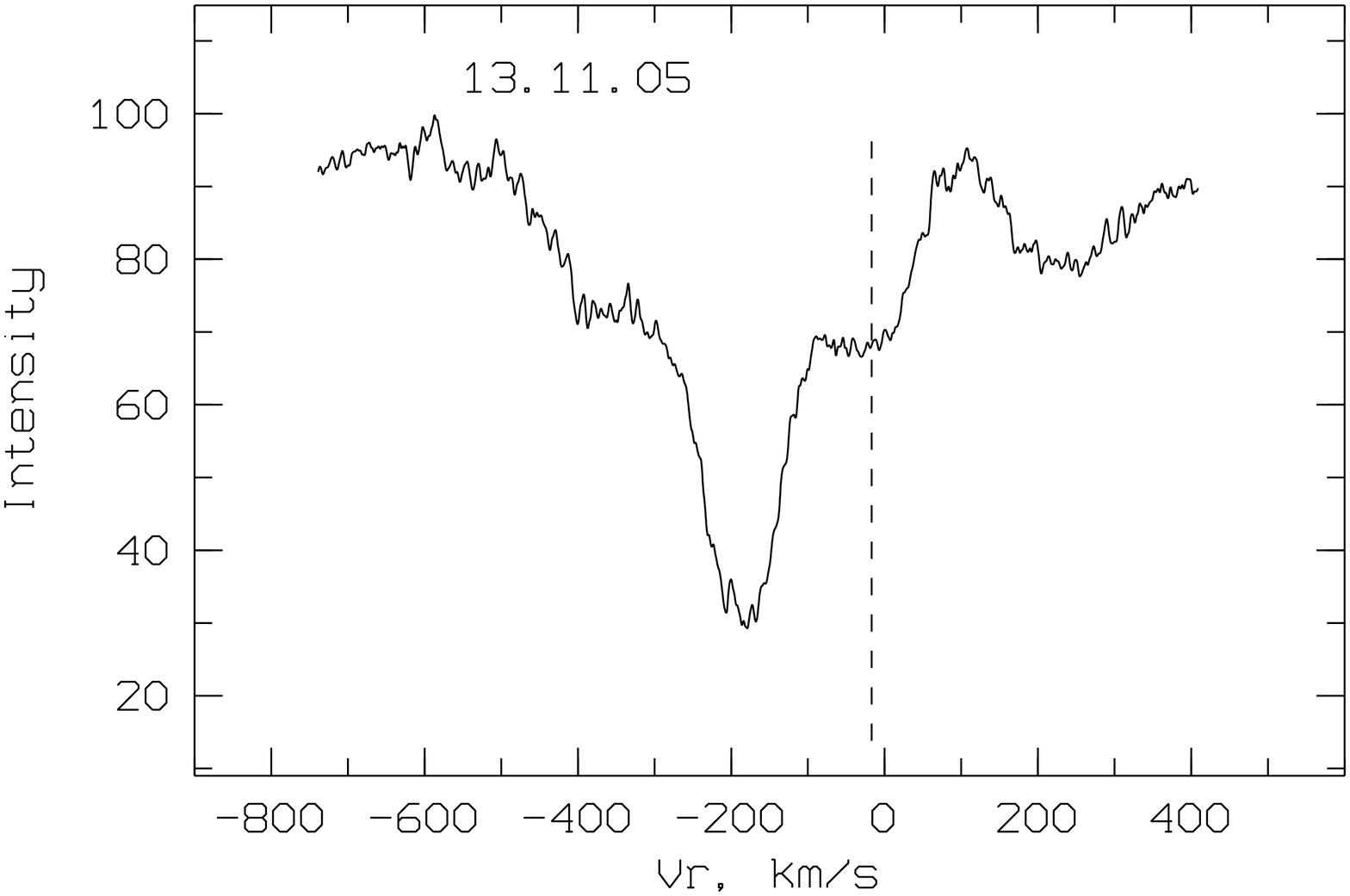}
\caption{H$\beta$-profile variability in  the spectra of  V2324\,Cyg.
        The dashed line shows the mean velocity averaged over photospheric
	metal lines.}
\label{Hbeta}
\end{figure*}
}

\clearpage
\newpage

\clearpage
\newpage

\newpage
\begin{table}[tbp]
\label{observ}
\caption{Log of observations of V2324\,Cyg made at the 6-m telescope}
\bigskip
\begin{tabular}{c|c|c|c|c}
\hline
Date  & UT &  $\Delta T$, s & $\Delta \lambda$,\,\AA &Spectrograph  \\ 
\hline
12.12.95&  16:24  & 6400  &  4720--6857  & Lynx  \\ 
12.06.01&  23:50  & 6000  &  4546--7880  & PFES  \\ 
14.04.03&  23:14  & 16200 &  5275--6760  & NES   \\ 
15.08.03&  21:26  & 10800 &  5275--6760  & NES   \\ 
20.08.03&  23:54  & 9600  &  4519--5997  & NES   \\ 
19.09.05&  16:29  & 12000 &  5275--6760  & NES   \\ 
13.11.05&  16:29  & 10800 &  4558--6010  & NES   \\ 
14.11.05&  16:40  & 10800 &  5275--6760  & NES   \\ 
01.09.06&  20:41  & 2400  &  5275--6760  & NES   \\
\hline
\end{tabular}
\label{observ}
\end{table}

\clearpage
\newpage

\begin{center}
\tablecaption{Residual intensities r and heliocentric radial velocities
              ${\rm V_r}$ for selected lines in the spectrum of V2324Cyg.
	      The colon indicates uncertain values}
\tablehead{\hline Line  & $\lambda$, \AA{} & $r$ & $V_r$, km/s & Date \\}
\tabletail{\hline \rule{0pt}{5pt}  1& 2  & 3 & 4 & 5   \\ \hline}
\begin{supertabular}{l|l|c|c|c}
\hline
FeII(38)  &  4576.34 &     0.86  &     $-$15    &        \\
\hline
FeII(38)  &  4583.84 &     0.83: &     $-$18    &        \\
\hline
CaI(23)   &  4585.87 &     0.88  &     $-$15    &        \\
\hline
CrII(44)  &  4588.20 &     0.89  &     $-$20    &    \\
\hline
TiII(50)  &  4589.95 &     0.88  &     $-$16     &    \\
\hline
FeI(39)   &  4602.94 &     0.90  &      $-$  &        \\
\hline
FeI(554)  &  4607.65 &     0.92: &      $-$  &    \\
\hline
FeI(554)  &  4613.28:&     0.88  &      $-$  &    \\
CrI(21)   &      &           &       &    \\
\hline
FeII(37)  &  4629.33 &     0.85  &     $-$15:    &    \\
FeI(115)  &      &           &       &    \\
\hline
CrII(44)  &  4634.07 &     0.89  &     $-$15     &    \\
\hline
FeI(820)  &  4643.47 &     0.94  &     $-$   &    \\
\hline
FeI(38)   &  4654.56:&     0.87  &     $-$18:    &    \\
FeI(554)  &      &           &       &    \\
\hline
FeII(25)  &          &           &       &    \\
ScII(24)  &  4670.30:&     0.86  &      $-$  &    \\
\hline
FeI(821)  &  4678.85 &     0.87  &      $-$  &    \\
\hline
MgI(11)   &  4702.99 &     0.82  &     $-$15     &        \\
\hline
NiI(98)   &  4714.42 &     0.87  &     $-$17     &    \\
\hline
FeII(43)  &  4731.47 &     0.86  &      $-$  &    \\
\hline
FeI(38)   &  4733.60 &     0.92  &      $-$  &    \\
\hline
FeI(554)  &  4736.78 &     0.86  &     $-$16     &    \\
\hline
TiII(92)  &  4778.99 &     0.85  &     $-$18     &    \\
\hline
MnI(16)   &  4783.42 &     0.87  &     $-$16     &    \\
TiII(92)  &  4805.09 &     0.88  &     $-$17:    &    \\
\hline
ZnI(2)    &  4810.54 &     0.95: &     $-$16:    &        \\
\hline
MnI(16)   &  4823.52 &     0.79  &     $-$   &    \\
CrII(30)  &  4824.14 &           &       &    \\
\hline
NiI(131)  &  4829.03 &     0.93: &      $-$  &    \\
CrI(31)   &  4829.37 &           &       &    \\
\hline
CrII(30)  &  4836.23 &     0.90  &     $-$   &    \\
\hline
CrII(30)  &  4848.25 &     0.84  &     $-$   &    \\
\hline
H$\beta$  &  4861.33 &     0.20: &    $-$138:   & 12.12.95 \\
          &          &     0.15  &    $-$145    & 12.06.01 \\
          &          &     0.19  &    $-$226    & 20.08.03 \\
          &          &     0.30  &    $-$182    & 13.11.05 \\
\hline
FeI(318)  &  4871.70:&     0.74  &      $-$  &    \\
\hline
FeI(318)  &  4891.10:&     0.75  &      $-$  &    \\
\hline
BaII(3)   &      &           &       &    \\
YII(22)   &  4900.10:&     0.86  &     $-$18:    &    \\
\hline
FeI(318)  &  4903.31 &     0.87  &      $-$  &    \\
\hline
FeI(318)  &  4918.90:&     0.81  &      $-$  &    \\
\hline
FeI(318)  &  4920.50 &     0.80  &     $-$16     &    \\
\hline
FeII(42)  &  4923.92 &     0.88  &      $-$  &    \\
FeI(114)  &  4924.77 &           &       &    \\
\hline
FeI(1065) &      &           &       &    \\
BaII(1)   &  4934.00:&     0.78  &      $-$  &    \\
\hline
FeI(687)  &  4950.11 &     0.94  &     $-$20:    &    \\
\hline
FeI(318)  &  4957.50:&     0.73  &      $-$  &    \\
\hline
FeI(687)  &  4966.09 &     0.91  &     $-$20:    &    \\
\hline
FeI(984)  &  4973.11 &     0.95  &     $-$16:    &    \\
\hline
FeI(1066) &  4988.98:&     0.94: &      $-$  &    \\
\hline
NiI(111)  &  5017.58 &           &       &    \\
FeII(42)  &  5018.40:&     0.85  &      $-$  &    \\
\hline
NiI(38)   &  5020.03 &     0.90: &      $-$  &    \\
\hline
ScII(23)  &  5031.02 &     0.89  &     $-$15     &    \\
\hline
FeI(383)  &  5068.80:&     0.88  &      $-$  &    \\
\hline
FeI(1094) &  5074.75 &     0.88  &     $-$16     &    \\
\hline
YII(20)   &  5087.42 &     0.90  &     $-$15     &    \\
\hline
FeI(1090) &  5090.78 &     0.93  &     $-$18     &    \\
\hline
FeI(16)   &  5107.55:&     0.87  &      $-$  &    \\
FeI(36)   &      &           &       &    \\
\hline
NiI(177)  &  5115.40 &     0.94: &      $-$  &        \\
\hline
TiII(86)  &  5129.16:&     0.87  &     $-$17:    &    \\
\hline
FeI(1092) &  5133.69 &     0.88  &     $-$20     &    \\
\hline
FeI(383)  &  5139.37:&     0.82  &     $-$17:    &    \\
\hline
TiII(70)  &  5154.07:&     0.86: &      $-$  &    \\
\hline
FeI(1089) &  5162.27 &     0.89  &     $-$21:    &    \\
\hline
FeI(1)    &  5166.28 &           &       &    \\
MgI(2)    &  5167.32:&     0.83: &      $-$  &    \\
FeI(37)   &  5167.49 &           &       &    \\
\hline
FeI(1)    &  5168.90 &           &       &        \\
FeII(42)  &  5169.03 &     0.89  &     $-$25:    &    \\
\hline
MgI(2)    &  5172.68 &     0.85  &     $-$25:    &    \\
\hline
MgI(2)    &  5183.61 &     0.83  &     $-$13     &    \\
\hline
TiII(86)  &  5185.91 &     0.90  &     $-$18     &    \\
\hline
TiII(70)  &  5188.69 &     0.83  &     $-$18:    &    \\
CaI(49)   &  5188.84 &           &       &    \\
\hline
FeII(49)  &  5197.58 &     0.87  &     $-$20:    &    \\
FeI(66)   &  5198.72 &           &       &    \\
\hline
FeI(66)   &  5202.34 &     0.91  &     $-$17     &    \\
\hline
CrI(7)    &  5208.44 &     0.84  &      $-$  &    \\
FeI(553)  &  5208.60 &           &       &    \\
\hline
FeI(553)  &  5229.85 &     0.91  &     $-$19     &    \\
\hline
CrII(43)  &  5232.50 &           &       &    \\
FeI(383)  &  5232.94 &     0.85  &      $-$  &    \\
\hline
CrII(43)  &  5237.32 &     0.89  &     $-$19     &    \\
\hline
ScII(26)  &  5239.82 &     0.93  &     $-$17     &    \\
FeI(1)    &  5247.06 &           &       &    \\
CrI(18)   &  5247.57 &     0.93: &      $-$  &    \\
\hline
FeI(1)    &  5250.22 &           &       &        \\
FeI(66)   &  5250.65 &     0.91: &      $-$  &    \\
\hline
FeI(553)  &  5273.28:&     0.88: &      $-$  &    \\
FeI(114)  &          &           &       &    \\
\hline
FeI(383)  &  5281.79 &     0.89  &      $-$  &    \\
\hline
FeI(929)  &  5288.53 &     0.96: &     $-$19:    &    \\
\hline
FeI(553)  &  5302.31 &     0.91  &     $-$18     &    \\
\hline
CrII(43)  &  5310.69 &     0.96  &     $-$15:    &    \\
\hline
CrII(43)  &  5313.58 &     0.94  &     $-$17     &    \\
\hline
FeII(49)  &  5316.66:&     0.87  &     $-$16:    &        \\
FeII(48)  &      &           &       &    \\
\hline
FeI(553)  &  5324.18 &     0.87  &      $-$  &    \\
\hline
CrI(18)   &  5345.80 &     0.93: &      $-$  &    \\
CrII(24)  &  5346.08 &           &       &    \\
\hline
FeII(48)  &  5362.86 &     0.90  &     $-$18     &    \\
\hline
FeI(1146) &  5364.87 &     0.89: &      $-$  &    \\
FeI(786)  &  5365.40 &           &       &    \\
\hline
FeI(1146) &  5367.47 &     0.90  &     $-$16     &    \\
\hline
FeI(1146) &  5383.37 &     0.89  &     $-$17     &    \\
\hline
FeI(553)  &  5393.17 &     0.91  &     $-$18     &    \\
\hline
FeI(15)   &  5397.13 &     0.89  &     $-$16:    &    \\
FeI(1145) &  5398.29 &           &       &    \\
\hline
FeI(1145) &  5400.51:&     0.93: &      $-$  &    \\
\hline
FeII(48)  &  5414.07 &           &       &    \\
FeI(1165) &  5415.20 &     0.89  &      $-$  &    \\
\hline
TiII(69)  &  5418.77 &     0.92  &     $-$16     &    \\
\hline
FeI(1146) &  5424.07 &     0.86  &      $-$  &    \\
FeII(49)  &  5425.25 &           &       &    \\
\hline
FeI(15)   &  5429.70 &     0.86  &     $-$16     &        \\
\hline
FeI(15)   &  5434.52 &     0.89  &     $-$16     &    \\
\hline
TiI(3)    &  5446.59 &           &       &    \\
FeI(15)   &  5446.92:&     0.83: &     $-$17:    &    \\
\hline
FeI(1145) &          &           &       &    \\
FeI(15)   &  5455.56:&     0.83  &     $-$17:    &    \\
\hline
FeI(15)   &  5497.52 &     0.91  &     $-$17     &    \\
FeI(15)   &  5506.79 &     0.91  &      $-$  &    \\
\hline
DIB       &  5512.64 &     0.92  &      $-$  &    \\
CaI(48)   &  5512.99 &           &       &    \\
\hline
ScII(31)  &  5526.81 &     0.91: &      $-$  &    \\
\hline
MgI(9)    &  5528.41 &     0.87  &      $-$  &    \\
\hline
FeII(55)  &  5534.84 &     0.87: &      $-$  &    \\
FeI(626)  &  5535.42 &           &       &    \\
\hline
FeI(1183) &  5554.90 &     0.94  &      $-$  &    \\
\hline
FeI(1183) &  5565.71 &     0.95  &     $-$15     &    \\
\hline
FeI(686)  &  5569.62 &     0.93  &     $-$15     &    \\
\hline
FeI(686)  &  5572.86:&     0.90  &     $-$16:    &    \\
\hline
FeI(686)  &  5576.10 &     0.94  &     $-$18:    &        \\
\hline
CaI(21)   &  5581.98 &     0.93  &     $-$17     &    \\
\hline
CaI(21)   &  5594.48:&     0.88  &     $-$18:    &    \\
FeI(1182) &  5594.66 &           &       &    \\
\hline
CaI(21)   &  5601.28 &     0.92: &      $-$  &    \\
\hline
CaI(21)   &          &           &       &    \\
FeI(686)  &  5602.90:&     0.88  &      $-$  &    \\
\hline
FeI(209)  &  5615.31 &           &       &    \\
FeI(686)  &  5615.60:&     0.86  &     $-$17:    &    \\
\hline
FeI(1314) &  5633.95 &     0.96: &     $-$15:    &    \\
\hline
FeI(1183) &  5679.03 &     0.96  &     $-$18     &    \\
\hline
NaI(6)    &  5688.21 &     0.93: &      $-$  &    \\
\hline
NiI(231)  &  5715.09 &     0.97: &      $-$  &    \\
\hline
FeI(1107) &  5717.84 &     0.97: &     $-$16:    &    \\
\hline
FeI(1087) &  5731.77 &     0.95: &     $-$15:    &    \\
\hline
FeI(1107) &  5763.00:&     0.92: &     $-$17:    &    \\
SiI(17)   &  5772.15 &     0.96  &     $-$16     &    \\
\hline
FeI(1087) &  5775.08 &     0.98: &      $-$     &     \\
\hline
DIB       &  5780.37 &     0.90  &      $-$7:    &        \\
\hline
DIB       &  5796.96 &     0.87  &      $-$8     &    \\
\hline
FeI(1179) &  5816.38 &     0.95: &     $-$16:    &    \\
\hline
DIB       &  5849.80 &     0.96  &      $-$8     &    \\
\hline
BaII(2)   &  5853.68 &     0.92  &     $-$16     &    \\
\hline
CaI(47)   &  5857.45 &     0.91  &      $-$  &    \\
NiI(228)  &  5857.75 &           &       &    \\
\hline
FeI(1181) &  5859.59 &     0.95: &      $-$  &    \\
FeI(1180) &  5862.36 &     0.95  &     $-$15:    &           \\
\hline
NaI(1)    &  5889.95 & 0.82/1.17 & $-$140/2:     & 12.12.95  \\
          &          & 0.90/1.10 & $-$160/$-$8:  & 12.06.01  \\
          &          & 0.90/1.22 & $-$220/$-$21: & 14.04.03  \\
          &          & 0.96/1.23 & $-$207/$-$18: & 15.08.03  \\
          &          & 0.84/1.20 & $-$210/$-$22: & 20.08.03  \\
          &          & 0.81/1.24 & $-$181/$-$25: & 19.09.05  \\
          &          & 0.90/1.20 & $-$188/$-$26  & 14.11.05  \\
          &          & 0.90/1.20 & $-$200/$-$16: &  1.09.06  \\
\hline
NaI(1)\,I.S.&        &     0.07  &     $-$13     &       \\
\hline
NaI(1)    &  5895.92 & 0.88/1.09 & $-$137/$-$2:  & 12.12.95  \\
          &          & 0.93/1.08 & $-$150/$-$15: & 12.06.01  \\
          &          & 1.02/1.19 & $-$169/$-$20: & 14.04.03  \\
          &          & 1.05/1.18 & $-$172/$-$37: & 15.08.03  \\
          &          & 0.98/1.16:& $-$180/$-$37: & 20.08.03  \\
          &          & 0.90/1.17 & $-$164/$-$20: & 19.09.05  \\
          &          & 0.94/1.12 & $-$168/$-$18  & 14.11.05  \\
          &          & 0.95/1.12 & $-$150/$-$37: &  1.09.06  \\
      &      &       &           &       \\
NaI(1)\,I.S.&        &     0.07  &     $-$13     &       \\
\hline
FeI(1180) &  5914.17:&     0.94: &     $-$16:    &       \\
FeI(1181) &      &           &           &       \\
\hline
FeI(1260) &  5987.07 &     0.96: &      $-$      &       \\
\hline
FeII(46)  &  5991.37 &     0.95  &     $-$17:    &       \\
\hline
FeI(959)  &  6003.02 &     0.97: &     $-$15:    &       \\
\hline
MnI(27)   &  6016.64 &     0.97  &     $-$15:    &       \\
\hline
FeI(1178) &  6024.07 &     0.95  &     $-$14:    &       \\
FeI       &  6042.10 &     0.96  &     $-$17:    &       \\
\hline
FeI(1259) &  6056.01 &     0.96  &     $-$16:    &       \\
\hline
FeI(207)  &  6065.49 &     0.93  &     $-$17:    &       \\
\hline
FeII(46)  &  6084.10 &     0.97: &      $-$  &       \\
\hline
FeI(1259) &  6102.18 &           &       &       \\
CaI(3)    &  6102.73:&     0.87  &     $-$16:    &       \\
FeI(1260) &  6103.19 &           &       &       \\
\hline
CaI(3)    &  6122.22 &     0.89  &     $-$17     &       \\
\hline
FeI(169)  &  6136.62 &           &       &       \\
FeI(62)   &  6137.00 &     0.87: &      $-$  &       \\
FeI(207)  &  6137.70 &           &       &       \\
\hline
BaII(2)   &  6141.72 &     0.86  &     $-$17     &       \\
\hline
FeII(74)  &  6147.74 &     0.93  &      $-$      &       \\
FeII(74)  &  6149.25 &           &       &       \\
\hline
CaI(20)   &  6161.29 &           &       &       \\
CaI(3)    &  6162.18:&     0.89  &     $-$18:    &       \\
\hline
DIB       &  6195.96 &     0.94  &     $-$11     &       \\
\hline
FeI(207)  &  6200.32 &     0.97  &      $-$  &       \\
\hline
DIB       &  6203.08 &     0.96  &     $-$12:    &       \\
\hline
FeI(207)  &  6230.73 &     0.92  &     $-$17     &       \\
\hline
FeI(62)   &  6265.14 &     0.96: &      $-$  &       \\
\hline
SiII(2)   &  6347.10 &     0.90  &     $-$14:    &       \\
\hline
SiII(2)   &  6371.36 &     0.93  &      $-$  &       \\
\hline
DIB       &  6379.29 &     0.91  &     $-$12     &       \\
\hline
FeI(168)  &  6393.61 &     0.93  &     $-$16     &       \\
\hline
FeI(816)  &  6400.01:&     0.91  &     $-$17:    &       \\
FeI(13)   &  6400.32 &           &       &       \\
FeI(816)  &  6411.66 &     0.95  &     $-$16:    &       \\
\hline
CaI(18)   &  6439.08 &     0.89  &     $-$17     &       \\
\hline
CaI(19)   &  6449.82 &     0.92  &     $-$18:    &       \\
\hline
CaI(19)   &  6455.60 &           &       &       \\
FeII(74)  &  6456.39:&     0.88  &     $-$16:    &       \\
\hline
CaI(18)   &  6462.61:&     0.90  &     $-$16:    &       \\
FeI(168)  &  6462.73 &           &       &       \\
\hline
FeI(268)  &  6546.25 &     0.91  &     $-$15     &       \\
\hline
H$\alpha$ &  6562.81 & 0.77/3.70:&  $-$180/31    & 12.12.95  \\
          &          & 0.40/3.00:&  $-$183/47    & 12.06.01  \\
          &          & 0.40/3.90 &  $-$247/45    & 14.04.03  \\
          &          & 0.29/3.90 &  $-$278/30    & 15.08.03  \\
          &          & 0.24/3.60 &  $-$260/42    & 19.09.05  \\
          &          & 0.25/3.60 &  $-$195/51    & 14.11.05  \\
          &          & 0.35/3.30 &  $-$280/40    &  1.09.06  \\
\hline
DIB       &  6613.56 &     0.87  &     $-$12:    &       \\
\hline
NiI(43)   &  6643.64 &     0.96  &     $-$13:    &       \\
\hline
DIB       &  6660.64 &     0.95: &     $-$10:    &       \\
\hline
FeI(268)  &  6678.00 &     0.94  &     $-$14:    &       \\
\hline
LiI(1)    &  6707.80:&     0.95  &     $-$15:    &       \\
\hline
CaI(32)   &  6717.69 &     0.94  &     $-$17     &       \\
\hline
CaI(30)   &  7148.16 &     0.89: &      $-$  &       \\
\hline
FeI(1077) &  7495.08 &     0.93  &     $-$15:    &       \\
\hline
FeI(1077) &  7511.03 &     0.94  &      $-$  &       \\
\hline
NiI(187)  &  7555.61 &     0.95  &      $-$  &       \\
\hline
FeI(1077) &  7568.91 &     0.94  &      $-$  &       \\
\hline
FeI(1306) &  7742.72 &     0.94  &      $-$  &       \\
\hline
OI(1)     &  7771.94 &     0.96: &     $-$16:    &       \\
\hline
OI(1)     &  7774.17 &     0.90: &      $-$  &       \\
OI(1)     &  7775.39 &           &       &       \\
\hline
FeI(1154) &  7780.57 &     0.93  &     $-$15:    &       \\
\hline
FeI(1154) &  7832.21 &     0.93: &      $-$  &       \\
\end{supertabular}
\label{lines}
\end{center}

\newpage
\begin{table}[tbp]
\caption{Uncertainty $\Delta\,\log\,\varepsilon(X)$ of elemental abundances in the atmosphere of
         V2324\,Cyg due to the errors of the parameters of the star
         atmosphere}
\medskip
\begin{tabular} {l|c|c|c}
\hline
X & $\Delta\,T_{eff}$& $\Delta\,\log\,g$  & $\Delta\,\xi_t$  \\
  & +200\,K          & +0.5             &  $-1.0$\,km/s     \\
\hline
He{\sc i}  & $-$0.57 & $-$0.04 & $+$0.08 \\
Li{\sc i}  & $+$0.07 & $-$0.06 & $+$0.02 \\
C{\sc i}   & $-$0.02 & $+$0.03 & $+$0.05 \\
O{\sc i}   & $-$0.10 & $+$0.09 & $+$0.01 \\
Na{\sc i}  & $+$0.02 & $-$0.07 & $+$0.04 \\
Mg{\sc i}  & $-$0.12 & $-$0.24 & $+$0.27 \\
Si{\sc i}  & $-$0.14 & $+$0.09 & $+$0.07 \\
S{\sc i}   & $+$0.02 & $-$0.02 & $+$0.03 \\
Ca{\sc i}  & $-$0.04 & $+$0.07 & $+$0.02 \\
Sc{\sc ii} & $-$0.03 & $+$0.13 & $+$0.04 \\
Ti{\sc ii} & $+$0.00 & $+$0.14 & $+$0.09 \\
Cr{\sc i}  & $+$0.06 & $-$0.05 & $+$0.09 \\
Cr{\sc ii} & $-$0.01 & $+$0.13 & $+$0.04 \\
Fe{\sc i}  & $+$0.03 & $-$0.05 & $+$0.14 \\
Fe{\sc ii} & $-$0.02 & $+$0.14 & $+$0.13 \\
Ni{\sc i}  & $+$0.04 & $-$0.06 & $+$0.05 \\
Zn{\sc i}  & $+$0.06 & $-$0.04 & $+$0.04 \\
Y{\sc ii}  & $+$0.03 & $+$0.27 & $+$0.07 \\
Ba{\sc ii} & $+$0.06 & $+$0.04 & $+$0.29 \\
\hline
\end{tabular}
\label{chem}
\end{table}

\newpage
\begin{table}[tbp]
\caption{Abundances $\varepsilon(X)$ of elements in the atmosphere of
         V2324\,Cyg. Here $\sigma$ is the standard deviation of the
	abundance and ``n'', the number of lines used in the computation.
	The computation was performed with the model-atmosphere parameters
        $T_{eff}$\,=7500\,K, $\log g$\,=\,2.0, and $\xi_t$\,=\,6.0\,km/s.
	The elemental abundances for the solar atmosphere are adopted
	from~[\cite{Grev}]}
\medskip
\begin{tabular}{l|  l | r  | r |  r  | r }
\hline
\multicolumn{2}{c|}{The Sun} &\multicolumn{4}{c}{V2324\,Cyg } \\  
\hline
X &$\varepsilon(X)$&$\varepsilon(X)$&$\sigma$&n& [X/Fe] \\ 
\hline
Li{\sc i}    & 3.25$^1$& $\ge$3.93&  &  1 &$\ge$+0.68\\ 
C{\sc i}     & 8.39  &   8.73 & 0.08 &  9 & +0.35    \\ 
O{\sc i}     & 8.66  &   8.53 & 0.28 &  3 &$-$0.12   \\ 
Na{\sc i}    & 6.17  &   7.20 & 0.22 &  3 & +1.04    \\ 
Mg{\sc i}    & 7.53  &   7.96 & 0.04 &  2 & +0.43    \\ 
Si{\sc ii}   & 7.51  &   7.22 & 0.09 &  5 &$-$0.28   \\ 
S{\sc i}     & 7.11  &   7.41 & 0.19 &  2 & +0.31    \\ 
Ca{\sc ii}   & 6.31  &   6.85 & 0.13 &  4 & +0.55    \\ 
Sc{\sc ii}   & 3.05  &   3.05 & 0.15 &  8 & +0.01    \\ 
Ti{\sc ii}   & 4.90  &   4.76 & 0.08 & 11 &$-$0.13   \\ 
Cr{\sc i}    & 5.64  &   5.39 & 0.10 &  5 &$-$0.24   \\ 
Cr{\sc ii}   &       &   5.31 & 0.06 & 19 &$-$0.32   \\ 
Fe{\sc i}    & 7.45  &   7.44 & 0.07 & 28 & +0.00    \\ 
Fe{\sc ii}   &       &   7.43 & 0.10 & 14 &$-$0.01   \\ 
Ni{\sc i}    & 6.23  &   6.23 & 0.10 &  8 & +0.01    \\ 
Zn{\sc i}    & 4.60  &   4.68 & 0.17 &  2 & +0.09    \\ 
Y{\sc ii}    & 2.21  &   2.18 & 0.27 &  2 &$-$0.02   \\ 
Ba{\sc ii}   & 2.17  &   2.64 & 0.09 &  4 & +0.46    \\ 
\hline
\multicolumn{6}{l}{\small Remark: 1---the value for meteorites [\cite{Grev}]} \\
\end{tabular}
\label{error}
\end{table}

\clearpage
\newpage
\begin{table}[tbp]
\caption{Equivalent widths $W_{\lambda}$ of interstellar NaI\,D lines
         and DIBs. Approximate estimates are given for the equivalent
	 widths of NaI\,D lines because of the complex shapes of their
	 profiles (see Fig.\,\ref{NaD})}
\bigskip
\begin{tabular}{c @{\qquad} r}
\hline
Spectral feature&  $W_{\lambda}$, \AA  \\ 
\hline
D2 5889  &$\approx$0.40   \\ 
D1 5895  &$\approx$0.37   \\ 
DIB 5512 &  0.08          \\ 
DIB 5780 &  0.23          \\ 
DIB 5797 &  0.13          \\ 
DIB 5850 &  0.03          \\ 
DIB 6196 &  0.04          \\ 
DIB 6203 &  0.04          \\ 
DIB 6379 &  0.07          \\ 
DIB 6614 &  0.12          \\ 
DIB 6661 &  0.03          \\
\hline
\end{tabular}
\label{DIB}
\end{table}

\end{document}